\documentclass[useAMS,usenatbib]{mn2e}

\usepackage{graphicx}
\usepackage{mathrsfs}
\usepackage{natbib}
\usepackage{amssymb}
\bibpunct[,]{(}{)}{;}{a}{}{,}

\def\totd{{\mathrm{d}}}
\def\sun{{\odot}}

\newcommand{\begit}{\begin{itemize}}
\newcommand{\enit}{\end{itemize}}
\newcommand{\begen}{\begin{enumerate}}
\newcommand{\enen}{\end{enumerate}}

\newcommand       \be           {\begin{equation}}
\newcommand       \ee           {\end{equation}}
\newcommand       \bea          {\begin{eqnarray}}
\newcommand       \eea          {\end{eqnarray}}

\newcommand{\beqa}{\begin{eqnarray}}
\newcommand{\eeqa}{\end{eqnarray}}

\title[Delayed Outflows from BH Accretion Tori]{Delayed outflows from black hole accretion 
tori following neutron star binary coalescence}
\author[R.~Fern\'andez and B.~D.~Metzger]{Rodrigo Fern\'andez$^1$ and 
Brian D. Metzger$^2$\\
$^1$ Institute for Advanced Study. Princeton, NJ 08540, USA.\\
$^2$ Department of Physics, Columbia University. New York, NY 10027, USA.}

\begin{document}

\date{Submitted to MNRAS}
\pagerange{\pageref{firstpage}--\pageref{lastpage}} 
\pubyear{2013}
\maketitle
\label{firstpage}

\begin{abstract}
Expulsion of neutron-rich matter following the merger of neutron star (NS)
binaries is crucial to the radioactively-powered electromagnetic counterparts
of these events and to their relevance as sources of $r$-process
nucleosynthesis.  Here we explore 
the long-term (viscous) evolution of remnant
black hole accretion disks formed in such mergers by means of
two-dimensional, time-dependent hydrodynamical simulations.  The evolution
of the electron fraction due to charged-current weak interactions is included, and
neutrino self-irradiation is modeled as a lightbulb that
accounts for the disk geometry and moderate optical depth effects.  Over several 
viscous times ($\sim 1$ s), 
a fraction $\sim 10\%$ of the initial disk mass
is ejected as a moderately neutron-rich wind ($Y_e\sim 0.2$)
powered by viscous heating and nuclear recombination, with
neutrino self-irradiation playing a sub-dominant role.
Although the properties of the outflow vary in time
and direction, their mean values in the heavy-element production region
are relatively robust to variations in the initial conditions of
the disk and the magnitude of its viscosity. 
The outflow is sufficiently
neutron-rich that most of the ejecta forms heavy $r$-process elements with mass
number $A \gtrsim 130$, thus representing a new astrophysical source of
$r$-process nucleosynthesis, distinct from that produced in the dynamical
ejecta.  Due to its moderately high entropy, disk outflows contain a small residual
fraction $\sim 1\%$ of helium, which could produce a unique
spectroscopic signature.  
\end{abstract}


\maketitle

\section{Introduction}

Coalescing neutron star (NS) and stellar mass black hole (BH) binaries are
among the most promising sources for detection with networks of ground-based
gravitational wave (GW) observatories, including Advanced LIGO and Virgo
(\citealt{Abramovici+92,Acernese+09,Caron+99,Abadie+10,Abadie+12}).  The GW
signal provides information on the properties of the merging binary (e.g.,
\citealt{Ajith+07,vanderSluys+08}); a potential probe of the equation of state
of dense matter (e.g., \citealt{Faber+02,Shibata05,Read+09,Bauswein+12}); and
even a test of General Relativity itself (e.g.~\citealt{Cornish+11}).    

To maximize the scientific benefit of a GW detection, it is important to
identify a coincident electromagnetic counterpart (e.g.~\citealt{Bloom+09}, \citealt{Metzger&Berger12}, \citealt{Nissanke+13}, \citealt{Bartos+12}).
The sky localization provided by initial Advanced LIGO/Virgo will be tens to
hundreds of square degrees (e.g.~\citealt{Fairhurst09}; \citealt{Wen&Chen10};
\citealt{Nissanke+11}), but a coincident X-ray, optical, or radio signal could
provide a much more accurate position, such that the host galaxy and redshift
of the merger could be identified.  Electromagnetic emission also indirectly
probes the (magneto-)hydrodynamics of the merger and its aftermath.

The BH created from a NS-NS merger, following a metastable hypermassive NS phase,
is typically surrounded by a thick
torus of mass $\sim 10^{-3}-0.1 M_{\odot}$ (\citealt{ruffert1999};
\citealt{Uryu+00}; \citealt{Rosswog&Liebendorfer03};
\citealt{Oechslin&Janka06}; \citealt{Chawla+10}; \citealt{Rezzolla+10};
\citealt{Hotokezaka+11}).  
A torus also forms following a BH-NS merger if the binary mass ratio is sufficiently low
that the NS is tidally disrupted outside the BH instead of being swallowed
whole 
(e.g.~\citealt{Faber+06}; \citealt{Duez+10};
\citealt{Foucart+11,Stephens+11,Foucart+12}; \citealt{East+12}). 
The short accretion
timescale $t_{\rm visc} \lesssim 1$ s, and large accretion rates $\dot{M}
\gtrsim M_{\odot}$ s$^{-1}$, of such remnant tori motivate the current
paradigm that NS-NS/NS-BH mergers are the central engines powering short
duration gamma-ray bursts (SGRBs) (\citealt{Paczynski86}; \citealt{Eichler+89}).

Despite being well-studied, SGRBs do not necessarily represent an ideal
counterpart for most GW-detected mergers
because SGRBs are difficult to localize on the sky and their observed rate is
relatively low ($\lesssim$ 1 yr$^{-1}$) within the sensitivity range of Advanced
LIGO/Virgo (\citealt{Metzger&Berger12}; \citealt{Chen&Holz12}).
These drawbacks have
motivated the study of more {\it isotropic} merger counterparts,
such as 
supernova-like optical transients powered by the
decay of radioactive elements synthesized in the merger ejecta
(`kilonovae', \citealt{Li&Paczynski98}; \citealt{Kulkarni05}; \citealt{Metzger+10}). 
Merger simulations show that a modest quantity of highly neutron rich material (electron fraction $Y_{e}
\lesssim 0.1$) is ejected by tidal forces during the merger
(\citealt{Lattimer&Schramm74,Rosswog+99,Rosswog05,Chawla+10,Stephens+11,East+12,Rosswog12, 
Rosswog+13,bauswein2013}),
and recent nuclear reaction network calculations largely agree on the amount of
radioactive heating provided by such decaying $r$-process elements
(\citealt{Metzger+10}; \citealt{Roberts+11}; \citealt{Goriely+11}).  

One large uncertainty, however, is the opacity of the ejecta, because current
kilonova models predict a composition dominated by heavy $r$-process nuclei
(atomic mass number $A \gtrsim 130$).  Recent atomic structure calculations
show that the opacity of material containing even a small quantity of Lathanide elements ($A \sim
139-175$) could be orders of magnitude higher than that of Fe
\citep{Kasen+13}.  This implies that the
predicted transient peaks on longer timescales -- up to a $\sim$ week -- and at a
lower luminosity and frequency -- in the near infrared rather than optical
\citep{Barnes&Kasen13,tanaka2013} -- than in original models that assumed opacity similar
to Fe (e.g.~\citealt{Metzger+10}, \citealt{Piran+13}).  Given the sensitive dependence of the
signal on the Lathanide fraction, it is thus crucial to identify any
diversity in the nucleosynthetic composition of the ejecta.

Another uncertainty is the mass of the ejecta $M_{\rm ej}$.  Although the
quantity of matter expelled in dynamical `tidal tails' from the merger can be
large ($\gtrsim 10^{-2}M_{\odot}$) in some cases (e.g.~eccentric mergers;
\citealt{East+12}), it is often found to be significantly smaller ($\sim
10^{-4}-10^{-3}M_{\odot}$; \citealt{Hotokezaka+13}).

Dynamical expulsion is not the only source of ejecta, however.  Mass loss also
occurs in outflows from the accretion disk over longer, viscous timescales.
Initially, neutrinos radiated from the disk -- or from the central hypermassive
NS prior to its collapse -- can heat and drive an outflow from the surface of
the disk (a `neutrino-driven' wind; e.g.~\citealt{McLaughlin&Surman05};
\citealt{Metzger+08a}; \citealt{Surman+08}; \citealt{Dessart+09}; \citealt{Kizivat+10}; \citealt{Wanajo&Janka12}).  A
potentially even larger quantity of mass is lost from the disk at later times
due to powerful outflows driven by viscous heating and recombination of free
nuclei into $\alpha$-particles
(\citealt{Lee&RamirezRuiz07,Beloborodov08,Metzger+08b,Metzger+09a,Lee+09}).  

\citet[ hereafter MPQ09]{Metzger+09a} constructed a height-integrated,
time-dependent model of the viscous spreading of accretion disks formed from
NS-NS/NS-BH mergers, including the evolution of the midplane electron fraction
$Y_{e}$ due to weak interactions.  MPQ09 find that
as the disk spreads due to angular momentum transport 
and the temperature decreases, $e^{-}/e^{+}$ captures become slow compared to the
viscous evolution timescale.  Since these captures both cool the disk
and affect its composition, the disk becomes geometrically thick and $Y_{e}$ freezes out.
Soon after freeze-out, free neutrons and protons recombine into $\alpha$
particles, starting near the outer edge of the disk and moving inwards with time. 
MPQ09 estimated that $\sim 20-50\%$ of the initial disk mass is unbound 
by this energy deposition, with a range of electron fractions 
$Y_{e} \sim 0.1-0.4$ resulting from varying local conditions
during freeze-out (see also \citealt{Lee&RamirezRuiz07}; \citealt{Beloborodov08}).  

\citet{Lee+09} performed axisymmetric (2D) global simulations of long term disk evolution, which
confirmed this process of He `evaporation' explicitly.  However, since they assumed
$\beta$-equilibrium instead of explicitly following the weak interactions, 
they could not accurately determine the energy released by
$\alpha$-recombination or the electron fraction of the final ejecta.  
The electron fraction of disk outflows is critically important because it
controls whether the heavy elements synthesized in the outflow are dominated by
2nd/3rd peak $r$-process elements (if $Y_{e} \lesssim 0.3-0.4$); lighter neutron-rich
elements (if $0.3-0.4 \lesssim Y_{e} \lesssim 0.495$); or significant quantities of
$^{56}$Ni ($Y_{e} \gtrsim 0.495$) (e.g.~\citealt{Hoffman+97}).  Given the
potentially large difference in the optical opacity of Fe-group and $r$-process
elements, and the different geometry of tidal tails and disk winds, 
the resulting kilonova light curve, including contributions from both sources of
ejecta, could in principle be much more complex than previously anticipated (\citealt{Barnes&Kasen13,tanaka2013}). 

Beyond their implication for electromagnetic emission, disk outflows from
binary NS mergers also represent an essentially unexplored astrophysical site
for heavy element synthesis.  This issue is particularly important because the
astrophysical origin of $r$-process nuclei remains a mystery
\citep{Qian&Wasserburg07}, and binary NS mergers could be an important$-$and
possibly dominant$-$source (\citealt{Lattimer&Schramm74,
Freiburghaus+99,Korobkin+12}).  Nevertheless, with the exception of neutrino-driven disk outflows (e.g.~\citealt{Surman+08}; \citealt{Wanajo&Janka12}), to date most studies
of the $r$-process in such environments focus on the dynamical ejecta.

In this series of papers, we explore the long term evolution of the accretion
disks formed by NS-NS/NS-BH mergers by means of axisymmetric (2D) time-dependent
hydrodynamic simulations.  In paper I (this work) we explore the dynamics of the disk
evolution and the properties of the resulting disk outflows.  In separate
works, our results for the disk outflows will be used to make predictions for
the resulting transient electromagnetic signature and nucleosynthetic yields.   

Our models employ a finite-volume hydrodynamic method, a realistic equation of state,
and a complete implementation of charged-current weak interactions in 
the optically-thin regime. Approximations are made in order to capture the key
physical processes at a low computational cost: (1) the effects of general relativity
on the disk dynamics are approximated via a pseudo-Newtonian potential, (2)
neutrino self-irradiation is modeled as a `lightbulb' that
accounts for the basic disk geometry and includes corrections for moderate 
optical depth effects, and (3) angular momentum transport is approximated
by an anomalous shear-stress tensor with an $\alpha$-viscosity. These approximations
enable us to evolve accretion disks for a large number of orbits ($\sim 1000$) and
thus perform a parameter space study on the outflow properties which can guide
future, more detailed calculations.

The paper is organized as follows.  In $\S\ref{s:quantities}$ we briefly
summarize the properties of the initial accretion disk.  The numerical
implementation is described in $\S\ref{s:model}$.  Results are presented in
$\S\ref{s:results}$, followed by a discussion in $\S\ref{s:discussion}$. Our
conclusions are summarized in $\S\ref{s:conclusions}$. The appendices provide 
additional information on the neutrino treatment as well as general information 
on accretion regimes and nucleosynthesis in NS merger disks.

\section{Disk Properties}
\label{s:quantities}

The mass of the torus formed during a NS-NS merger can vary considerably,
$M_{\rm t}\sim 10^{-3}-0.1M_{\odot}$, depending on the binary mass ratio, the
assumed EOS, the eccentricity of the initial orbit, and whether the system 
undergoes a long-lived hypermassive NS phase prior to BH formation (see \citealt{Faber&Rasio12} for a recent review). 
The disk mass in a NS-BH merger also depends on the spin of the black hole, since spin controls
the location of the innermost circular orbit.  If the NS is tidally disrupted
outside the BH horizon, then the torus mass is relatively large $M_{\rm t}
\sim 0.1M_{\odot}$; otherwise, the NS is swallowed whole and little or no torus
forms (e.g.,~\citealt{Foucart12}).  Merger simulations find that the initial torus is distributed across a wide
range of radii, with most of the mass and angular momentum concentrated on a
radial scale $R_0 \sim 20-100$ km.  

The (Newtonian) orbital time at this radial scale is
\begin{equation}
\label{eq:tdyn}
t_{\rm orb} = 3.5\times 10^{-3}\left(\frac{R_0}{50\,\textrm{km}}\right)^{3/2}
\left( \frac{M_{\rm BH}}{3M_\sun}\right)^{-1/2}\textrm{ s},
\end{equation}
where $M_{\rm BH}$ is the BH mass.  The characteristic timescale for matter to accrete may be estimated by the viscous time
\begin{eqnarray}
\label{eq:tacc}
t_{\rm visc}& \simeq &\alpha^{-1}\left(\frac{R_{0}^{3}}{GM_{\rm BH}}\right)^{1/2}\left(\frac{H_{0}}{R_{0}}\right)^{-2}\nonumber \\
&\sim& 1{\rm\,s} \left(\frac{0.03}{\alpha}\right)\left(\frac{R_{0}}{50{\,\rm km}}\right)^{3/2}
\left(\frac{3M_\sun}{M_{\rm BH}}\right)^{1/2}
\left(\frac{H_0}{3R_{0}}\right)^2,\nonumber\\
\label{eq:tvisc}
\end{eqnarray}
where $H_0$ is the vertical scaleheight and $\alpha$ parametrizes the disk viscosity ($\S\ref{s:viscosity}$), 
resulting in a characteristic accretion rate
\begin{eqnarray}
\label{eq:mdot_out}
\dot{M}_{0}  &\sim& \frac{M_{\rm t}}{t_{\rm visc}} \nonumber \\
&\sim & 10^{-2}M_{\sun}{\,\rm s^{-1}}\left(\frac{\alpha}{0.03}\right)
\left(\frac{M_{\rm t}}{10^{-2}M_{\sun}}\right)\left(\frac{R_{0}}{50{\,\rm km}}\right)^{-3/2}\nonumber\\
& &\qquad\qquad\qquad\times \left(\frac{M_{\rm BH}}{3M_\sun}\right)^{1/2} \left(\frac{3H_0}{R_{0}}\right)^{2}.
\label{eq:Mdot0}
\end{eqnarray}

Due to the high densities and rapid evolution of the disk, photons are
not an important source of cooling on timescales of relevance.  However, the
temperature is sufficiently high that neutrinos are relevant, especially in the 
earliest phases of the disk evolution (e.g., \citealt{popham1999}; \citealt{Narayan+01};
\citealt{DiMatteo+02}; \citealt{Chen&Beloborodov07}; hereafter CB07).  A
straightforward calculation shows that a thin neutrino-cooled disk ($H \sim
0.1-0.3R_{0}$) obtains when the accretion rate at radius $R_{0}$ exceeds a
critical value $\dot{M}_{\nu}$ given by (see \citealt{Metzger+08a}, their eq.~[11]) 
\be
\dot{M}_{\nu} \approx 1.3\times 10^{-2}M_{\odot}{\rm s^{-1}}\left(\frac{R_{0}}{50{\rm km}}\right)^{5/6}\left(\frac{\alpha}{0.03}\right)^{5/3}\left(\frac{M_{\rm BH}}{3M_{\odot}}\right)^{1/2},\label{eq:Mdotnu}
\ee
where the dominant source of cooling is assumed to be charged-current weak interactions,
and general relativistic corrections have been neglected.  Unlike photon-cooled
disks, such as those in X-ray binaries or AGN, neutrino-cooled disks are the most radiatively efficient at small radii.

By equating $R_{0}$ in equation (\ref{eq:Mdotnu}) with the radius $R_{\rm
isco}$ of the innermost stable circular orbit of the BH, one obtains a critical
accretion rate $\dot{M}_{\rm ign}$ (the `ignition' rate) above which the disk
is thin near its innermost radii:
\be
\dot{M}_{\rm ign} \approx 6\times 10^{-2}[2\times 10^{-2}]\left(\frac{\alpha}{0.03}\right)^{5/3}\left(\frac{M_{\rm BH}}{3M_{\odot}}\right)^{4/3}M_{\odot}{\rm s^{-1}},
\label{eq:Mdotign}
\ee
where the prefactor was calculated by CB07 for a 3$M_{\odot}$ black hole of
spin $a = 0$[$a = 0.95$] using a steady-state accretion model in a Kerr metric.  

When the disk is neutrino-cooled ($\dot{M} > \dot{M}_{\nu}$) its internal
energy $e_{\rm int}$ is dominated by non-degenerate nucleons (CB07).  The
characteristic temperature of the torus midplane, which is attained on a
cooling timescale $t_{\rm cool} \sim t_{\rm visc}(H/R_{0})^{2}$, 
is thus approximately given by
\begin{equation}
\label{eq:Tvir_def}
T_{\rm vir} \simeq 7.6(1+Y_{e})\left(\frac{M_{\rm BH}}{3M_\sun}\right)\left(\frac{3H_{0}}{R_{0}}\right)^{2}\left(\frac{50\textrm{km}}{R_0}\right)\textrm{ MeV}, 
\end{equation}
where $Y_{e} = n_{\rm p}/(n_{\rm n}+n_{\rm p})$ is the electron fraction and
$n_{\rm p}$($n_{\rm n}$) are the number densities of protons(neutrons),
respectively.  The initial temperature of the disk is sufficiently high that
all nuclei are dissociated into free nucleons.  Alpha particles and heavier nuclei 
form only once the disk temperature decreases to $\lesssim$ 1 MeV. 

\section{Numerical Setup}
\label{s:model}

\subsection{Equations Solved and Numerical Method}
\label{s:equations}
\label{s:numerical}

We solve the equations of mass, momentum, energy, and lepton number 
conservation on an axisymmetric spherical polar grid $(r,\theta)$, with 
source terms due to gravity, shear viscosity, and optically-thin weak interactions
\begin{eqnarray}
\label{eq:mass_conservation}
\frac{\partial \rho}{\partial t} + \nabla \cdot (\rho\mathbf{v}_p) & = & 0\\ 
\label{eq:momentum_conservation}
\frac{\totd \mathbf{v}_p}{\totd t}  & = &
\mathbf{f}_{\rm c}-\frac{1}{\rho}\nabla p  -\nabla\Phi \\
\label{eq:angular_conservation}
\rho\frac{\totd \ell_z}{\totd t} & = & r\sin\theta\,(\nabla\cdot\mathbb{T})_\phi\\
\label{eq:energy_conservation}
\rho\frac{\totd e_{\rm int}}{\totd t} + p\nabla\cdot\mathbf{v}_p 
& = & \frac{1}{\rho\nu}\mathbb{T}:\mathbb{T} + \rho\dot{Q}_{\rm net}\\
\label{eq:lepton_conservation}
\frac{\totd Y_e}{\totd t} & = & \Gamma_{\rm net}.
\end{eqnarray}
Here $\rho$, $p$, $\mathbf{v}_p=v_r\hat r + v_\theta\hat\theta$, and $e_{\rm
int}$ denote the fluid density, pressure, poloidal velocity, and internal energy,
respectively. The Lagrangian differential operator is $\totd/\totd t\equiv \partial/\partial t +
\mathbf{v}_p\cdot\nabla$. 
The specific angular momentum along the symmetry axis is given by $\ell_z =
r\sin\theta v_\phi$, with $v_\phi$ the azimuthal velocity, and $\mathbf{f}_c$ is
the centrifugal force in the poloidal direction.

The pseudo-newtonian gravitational potential of \citet{paczynsky1980} is employed,
$\Phi = -GM_{\rm BH}/(r - R_{\rm S})$, with $R_{\rm S} = 2GM_{\rm BH}/c^{2}$ the Schwarzschild radius,
as applicable to a BH of moderate spin.\footnote{BHs formed from NS-NS mergers have typical 
spin parameter $a \sim 0.7-0.8$ (e.g.,~\citealt{Rezzolla+10})} The black hole mass
is kept constant in the simulation.
The neutrino source terms are $\dot{Q}_{\rm net}$
and $\Gamma_{\rm net}$ (see \S\ref{s:neutrinos}), and the viscous stress tensor is denoted
by $\mathbb{T}$ (see \S\ref{s:viscosity})

The \emph{Helmholtz} EOS \citep{timmes2000} is used to relate 
internal energy and pressure. 
The ion component is an ideal gas of neutrons, protons, and alpha particles, with relative
abundances that satisfy nuclear 
statistical equilibrium (NSE). 
Heavier nuclei are not included, since the energy
release in their formation is much smaller in comparison, and hence should have little
impact on the outflow dynamics\footnote{The formation of heavy nuclei would indeed occur at radii 
within our simulation grid, however, see $\S\ref{s:nucleosynthesis}$}. 
An additional (inert) hydrogen gas is used to populate the ambient
medium that initially surrounds the torus.
The zero point of energy is taken to be the pure nucleon state, with the internal energy becoming 
\begin{equation}
\label{eq:recombination_energy}
e_{\rm int} = e_{\rm int,0} - \frac{Q_\alpha}{m_\alpha}\,X_\alpha,
\end{equation}
where $e_{\rm int,0}$ is the specific internal energy provided by the \emph{Helmholtz} EOS, $X_\alpha$ is the mass fraction 
of alpha particles, $m_\alpha$ is the mass of an alpha particle, 
and $Q_\alpha\simeq 28.3$~MeV is the nuclear binding energy of an alpha particle (e.g., \citealt{audi2003}). 
In addition, the thermodynamic derivatives of the pressure, internal energy, and entropy of the ions 
acquire terms that originate in the dependence of the ion abundances on temperature and density.
If the temperature falls below $5\times 10^9$~K, or if the mass fraction of inert hydrogen
exceeds 1\%, the abundances are frozen. Any hydrogen mixed with torus material is added to the
proton mass fraction if the temperature is higher than $5\times 10^9$~K.

We use FLASH3.2 \citep{dubey2009} to evolve the system of
equations~(\ref{eq:mass_conservation})-(\ref{eq:lepton_conservation}) with the
dimensionally-split version of the Piecewise Parabolic Method (PPM,
\citealt{colella84}).  The public version of the code has been modified to
allow for a non-uniformly spaced grid in spherical polar coordinates, as
described in \citet{F12}.

The computational grid is logarithmically spaced in radius, covering the
range $2R_{\rm S}$ to $2\times 10^3R_{\rm S}$ in most models. In the polar direction, the
cell spacing is uniform in $\cos\theta$, covering the interval $[0,\pi]$.
We adopt this grid structure because we intend to cover a large dynamic range
in radius, and to concentrate the resolution near the midplane, given the quasi-spherical
character of the outflows.
Our standard resolution uses $64$ cells per decade in radius and $56$ cells
in the polar direction, yielding approximately square cells at the midplane $(\Delta r/r \simeq
\Delta \theta \simeq 2^\circ)$. The coarsest angular cell next to the polar axis
is $10.8^\circ$. One model is evolved at double the resolution in both radius
and in angle, to quantify convergence. 
 
The boundary conditions at the polar axis are reflecting in all variables.
At the inner and outer radial boundaries, we impose a zero-gradient boundary condition
for all variables. In addition, the radial velocity in the ghost cells of the radial
boundaries is set to zero if its value in the first active cell next to the boundary 
would imply mass entering the domain.

All energy and viscous source terms are set to zero when the density falls
below a fiducial value of $10$~g~cm$^{-3}$. This value is sufficiently low as
to not impact the outflows from the torus, 
yet sufficiently high to avoid requiring too small of a time step due to angular momentum transport in
(unimportant) low-density regions.

To prevent numerical problems in the funnel that develops around the rotation
axis, we impose a floor of density equal to $1$~g~cm$^{-3}$,
and a floor of temperature at $10^5$~K.

\subsection{Neutrino Treatment}
\label{s:microphysics}
\label{s:neutrinos}

The inner regions of the disk become opaque to neutrinos
above an accretion rate (Appendix~\ref{s:mdot_opaque})
\be \dot{M}_{\rm opaque} \simeq 
0.15M_{\odot}{\rm\,s^{-1}}\,\left(\frac{\alpha}{0.03}\right)^{4/5}
\left(\frac{M}{3M_{\odot}}\right)^{7/10}\left(\frac{R_{\rm isco}}{6R_{g}}\right)^{9/5}.
\label{eq:Mdotopaque_copy}
\ee
The condition $\dot{M}_{\rm opaque} \gtrsim \dot{M}_{0}$ (eq.~[\ref{eq:Mdot0}]), 
sets the minimum disk mass for optical thickness.
For slowly spinning BHs, this mass is $\sim 0.1M_\sun$.

Most of the disks we explore ($M_t \lesssim 0.1M_{\sun}$) are thus only marginally optically thick to neutrinos at
early times. Since the disk density decreases as accretion proceeds, full neutrino transparency
is always achieved. The models we explore undergo this transition before 
weak freezout occurs, as quantified by the magnitude of the optical depth, and hence we 
adopt an optically thin treatment of energy exchange with matter and 
regulation of the electron fraction in all cases. Finite optical depth corrections are included
to prevent excessive heating at early times in the few massive tori we evolve. 

Charged-current weak interactions are included as formulated by \citet{bruenn85} and
implemented by \citet{F12}. These reactions control the evolution of the electron fraction, and provide
the dominant energy exchange channel between neutrinos and matter. The relevant source terms are 
the net rate of change of the electron fraction
\begin{equation}
\label{eq:gamma_net}
\Gamma_{\rm net} = \Gamma_{e-} - \Gamma_{e+},
\end{equation}
where $\Gamma_{e-}$ and $\Gamma_{e+}$ are the rates of electron and positron creation per baryon, respectively,
and the net specific heating rate of the gas
\begin{equation}
\label{eq:Q_net}
\dot{Q}_{\rm net} = \mathcal{H}_{\nu_e} + \mathcal{H}_{\bar{\nu}_e} - \mathcal{C}_{\nu_e} -
                    \mathcal{C}_{\bar{\nu}_e} - \mathcal{C}_{\rm pairs},
\end{equation}
where $\mathcal{H}_{\nu_e}$ and $\mathcal{H}_{\bar{\nu}_e}$ is the heating by absorption of neutrinos and antineutrinos
on nucleons, respectively, and $\mathcal{C}_{\nu_e}$ and $\mathcal{C}_{\bar{\nu}_e}$ are the cooling rates from
electron and positron capture, respectively. Explicit forms for these terms are provided in Appendix~\ref{s:weak_rates_appendix}.
Additional neutrino energy losses from pair creation ($\mathcal{C}_{\rm pairs}$) are included via
the \citet{itoh1996} parameterization.\footnote{http://cococubed.asu.edu/code\_pages/codes.shtml}
Source terms are included via explicit operator split in the equations for energy and lepton number
conservation.

Self-irradiation of the disk is implemented via a lightbulb-type prescription that accounts for the 
disk geometry (Appendix~\ref{s:weak_rates_appendix}). All of the neutrino emission is
assumed to come from a ring of material on the midplane, with a radius $R_{\rm em}$ chosen
to be an average value weighted by the neutrino emissivity, 
\begin{equation}
\label{eq:R_em_def}
R_{\rm em} = \frac{\int\sin\theta\,\totd\theta\,\totd r\, r^3\rho(\mathcal{C}_{\nu_e} + \mathcal{C}_{\bar{\nu}_e})}
		  {\int\sin\theta\,\totd\theta\,\totd r\, r^2\rho(\mathcal{C}_{\nu_e} + \mathcal{C}_{\bar{\nu}_e})}.
\end{equation}
This approximation is justified by the highly peaked form of this emissivity (Appendix~\ref{s:weak_rates_appendix}), 
and the low optical depth of most
of the disk masses considered.
The neutrino luminosities of electron-type neutrinos $L_{\nu_e}$ and antineutrinos $L_{\bar\nu_e}$ 
used in this self-irradiation treatment are calculated at the end of the previous time step by integrating 
the corresponding emissivities over the whole torus. 
Spectra are assumed to follow a Fermi-Dirac distribution with zero chemical potential,
with a temperature equal to the gas temperature in an angle-averaged radial shell ($10\%$ spread) 
around the emission radius $R_{\rm em}$.
Self-irradiation is included to explore its influence on
the ejecta composition and its relative energetic contribution to the total mass loss.
Results are not expected to be very sensitive to the particular choices of luminosities
and spectra, because the process is energetically subdominant (e.g., \citealt{ruffert1999}, MPQ09).

To explore more massive tori, finite optical depth effects are 
approximated by a modified version of the \citet{lee2005} treatment. At each time step, four 
optical depth functions for each neutrino species are computed from 
the neutrino emission peak,
\begin{eqnarray}
\label{eq:tau_up}
\tau_i^{\rm up}(\theta)  & = & \int_0^{\theta} \kappa_i\, R_{\rm em}\totd \theta^\prime,
				\qquad  \theta\in[0,\pi/2]\textrm{, }r=R_{\rm em }\\
\label{eq:tau_dn}
\tau_i^{\rm dn}(\theta)  & = & \int_\theta^\pi \kappa_i\, R_{\rm em}\totd \theta^\prime,
				\qquad  \theta\in[\pi/2,\pi]\textrm{, }r=R_{\rm em}\\
\label{eq:tau_out}
\tau_i^{\rm out}(r)      & = & \int_{r}^{r_{\rm out}} \kappa_i\,\totd r^\prime,
				\qquad  r\in[R_{\rm em},r_{\rm out}]\textrm{, }\theta=\frac{\pi}{2}\\
\label{eq:tau_in}
\tau_i^{\rm in}(r)       & = & \int_{r_{\rm in}}^{r} \kappa_i\, \totd r^\prime,
				\quad\qquad  r\in[r_{\rm in},R_{\rm em}]\textrm{, }\theta=\frac{\pi}{2},
\end{eqnarray}
corresponding to the four coordinate directions. Given that most of the torus mass lies within $\sim 45^\circ$ of
the equator, we use angular integration to estimate the vertical optical depth.
For fast computation, the absorption coefficients for this optical depth 
functions\footnote{Not to be confused with the absorption coefficients
used in the actual neutrino absorption rates (Appendix~\ref{s:weak_rates_appendix}).} are approximated as in \citet{janka2001}
\begin{eqnarray}
\kappa_{\nu_e} & \simeq & 2.2\times 10^{-7}\,\rho_{10}T^2_{\nu_e,4}\,\sqrt{X_p(2X_n+X_p)}\textrm{ cm}^{-1},\\
\kappa_{\bar\nu_e} & \simeq & 2.2\times 10^{-7}\,\rho_{10}T^2_{\nu_e,4}\,\sqrt{X_p(2X_p+X_n)}\textrm{ cm}^{-1}.
\end{eqnarray}
The neutrino emissivities $\mathcal{C}_{\nu_e}$ and $\mathcal{C}_{\bar{\nu}_e}$ are then suppressed 
everywhere by a factor $\exp{(-\tau_{i,\rm peak})}$, where
\begin{equation}
\label{eq:tau_peak_def}
\tau_{i,\rm peak} = \min[\tau_i^{\rm up}(\pi/2), \tau_i^{\rm dn}(\pi/2),\tau_i^{\rm out}(R_{\rm em}), \tau_i^{\rm in}(R_{\rm em})]. 
\end{equation}
This prescription accounts for the fact that 
most neutrinos will escape along the path with the least optical depth, which in most cases is the
vertical direction. In contrast to \citet{lee2005}, we ignore the neutrino pressure contribution,
which is a small correction (our tori are not radiation-dominated while optically thick).

The neutrino luminosity used for self-irradiation, first computed from the attenuated emissivities, 
is further suppressed by a local factor $\exp({-\tau_{i,\rm irr}})$, where
\begin{eqnarray}
\label{eq:tau_irr}
\tau_{i,\rm irr}(r,\theta) & = & \left[\tau_i^{x}(R_{\rm em})-\tau_i^{x}(r)\right]\nonumber\\
 && +  \frac{\tau_i^x(r)}{\tau_i^x(R_{\rm em})}\left[\tau_i^{y}(\pi/2)-\tau_i^{y}(\theta)\right].
\end{eqnarray}
The superscripts $x$ and $y$ are the radial and angular optical depths for the appropriate quadrant, 
respectively (eqns~[\ref{eq:tau_up}]-[\ref{eq:tau_in}]). This prescription attempts to capture the attenuation
of the neutrino flux in regions far from the emission peak, and smoothly approaches the semi-transparent lightbulb 
limit for decreasing torus density. It is a coarse approximation, however, since it does not account for 
attenuation along lateral directions, and assumes that the vertical density and abundance dependence
differs only by a constant scaling factor at different radii.

In optically thick disks, the rate of change of the electron fraction (eq.~[\ref{eq:gamma_net}]) is
affected indirectly via the suppression of the self-irradiation luminosities. The portion 
due to the local neutrino emissivities is not modified, however, since its contribution will 
still vanish in beta equilibrium.

\subsection{Angular Momentum Transport}
\label{s:viscosity}

Angular momentum transport is mediated by a viscous stress tensor $\mathbb{T}$ for
which only the azimuthal components are non-zero (e.g., \citealt{stone1999}),
\begin{eqnarray}
\label{eq:trphi_def}
T_{r\phi}      & = & \rho \nu\,\frac{r}{\sin\theta}\frac{\partial}{\partial r}\left(\frac{\ell_z}{r^2} \right)\\
T_{\theta\phi} & = & \rho \nu\,\frac{\sin\theta}{r^2}\frac{\partial}{\partial\theta}\left(\frac{\ell_z}{\sin^2\theta} \right).
\end{eqnarray}
Including all components of the stress would 
suppress convection in the poloidal direction (e.g.,
\citealt{igumenshchev1999}).  Our simulations mimic turbulent angular momentum 
transport via thermally-driven convection.  
Given the low mass of the torus relative to that of the central BH,
and its relatively thick scaleheight, 
we consider our neglect of self gravity as a good approximation. 

To connect with previous calculations, we employ a \citet{shakura1973} parameterization
of the kinematic viscosity coefficient
\begin{eqnarray}
\label{eq:viscosity_alpha}
\nu_\alpha & = & \alpha \frac{c_i^2}{\Omega_K}\nonumber\\
             & = & \alpha \left( \frac{c_i^2}{GM_{\rm BH}/r}\right) \sqrt{GM_{\rm BH}R_0}
		   \left(1-\frac{R_{\rm S}}{r} \right)\left( \frac{r}{R_0}\right)^{1/2}
\end{eqnarray} 
where $\Omega_K$ is the Keplerian frequency of the \citet{paczynsky1980} 
potential and $c_i^2 = p /\rho$ is the isothermal sound speed.

\begin{table*}
\centering
\begin{minipage}{18cm}
\caption{Models Evolved and Outflow Properties\label{t:models}\label{t:results}. 
Columns from left to right are: initial torus mass, black hole mass, 
initial torus radius (eq.~[\ref{eq:pp_general}]), initial electron fraction, 
initial entropy, viscosity coefficient (eq.~[\ref{eq:viscosity_alpha}]), spatial resolution (std=standard,
hi=high; \S\ref{s:numerical}), inclusion of recombination energy (eq.~[\ref{eq:recombination_energy}]), inclusion
of self-irradiation (\S\ref{s:neutrinos}), mass-flux-weighted values of the electron fraction, entropy,  
and expansion time $t_{\rm } = r/\bar{v}_r$ at a radius where $\bar{T}\sim 5\times 10^9$~K 
and within $\pm 60^\circ$ of the midplane (eq.~[\ref{eq:mean_value}]), 
ratio of the total mass with positive energy ejected beyond $r=10^9$~cm to the
total mass accreted through the ISCO radius, each in units of the initial torus mass, 
and mass-flux-weighted density (in units of $100$~g~cm$^{-3}$) and velocity (in units of $10,000$~km~s$^{-1}$) 
at $r=10^9$~cm and within $\pm 60^\circ$ of the midplane.}
\begin{tabular}{lccccccccccccccc}
{Model}&
{$M_{\rm t0}$} &
{$M_{\rm BH}$} &
{$R_0$} &
{$Y_{e0}$} &
{$s_0$} & 
{$\alpha$} &
{Res.} &
{Rec.} &
{Irr.} &
{$\bar{Y}_e$} &
{$\bar{s}$} &
{$t_{\rm exp}$} &
{$M_{\rm ej}/M_{\rm acc}$} &
{$\bar{\rho}_2$}  &
{$\bar{v}_{r,9}$} \\
{ } & \multicolumn{2}{c}{($M_\sun$)} & {(km)} & {} & {($k_{B}$/b)} & 
      {} & {} & {} & {} & {} & {($k_{B}$/b)} & {(ms)} & 
      {($M_{\rm t0}/M_{\rm t0}$)} & {} & {}\\
\hline
S-def       & 0.03 & 3   &   50  & 0.10  & 8   & 0.03  & std & yes & yes & 0.16 & 16  & 95   &  0.10/0.90 & 1.6 & 2.2 \\
\noalign{\smallskip}                                                                                                 
S-m0.01     & 0.01 & 3   &   50  & 0.10  & 8   & 0.03  & std & yes & yes & 0.15 & 18  & 68   &  0.10/0.90 & 0.5 & 2.2 \\
S-m0.10     & 0.10 &     &       &       &     &       &     &     &     & 0.17 & 17  & 119  &  0.11/0.89 & 6.3 & 2.3 \\
S-r75       & 0.03 & 3   &   75  &       &     &       &     &     &     & 0.16 & 16  & 126  &  0.22/0.78 & 3.6 & 2.2 \\
S-M10       &      & 10  &  150  &       &     &       &     &     &     & 0.18 & 17  & 77   &  0.04/0.92 & 0.8 & 1.8 \\
S-y0.05     &      & 3   &   50  & 0.05  &     &       &     &     &     & 0.15 & 16  & 93   &  0.10/0.90 & 1.5 & 2.2 \\
S-y0.15     &      &     &       & 0.15  &     &       &     &     &     & 0.17 & 16  & 92   &  0.10/0.90 & 1.6 & 2.3 \\
S-s6        &      &     &       & 0.10  & 6   &       &     &     &     & 0.14 & 17  & 86   &  0.08/0.91 & 1.2 & 1.9 \\
S-s10       &      &     &       &       & 10  &       &     &     &     & 0.19 & 16  & 87   &  0.12/0.88 & 1.9 & 2.6\\
S-v0.01     &      &     &       &       & 8   & 0.01  &     &     &     & 0.20 & 18  & 193  &  0.11/0.89 & 1.5 & 2.5 \\
S-v0.10     &      &     &       &       &     & 0.10  &     &     &     & 0.14 & 19  & 38   &  0.13/0.87 & 2.2 & 2.4 \\
S-hr        &      &     &       &       &     & 0.03  & hi  &     &     & 0.15 & 16  & 94   &  0.10/0.90 & 1.5 & 2.2 \\
\noalign{\smallskip}                                                                                                 
P-rec       & 0.03 & 3   &   50  & 0.10  & 8   & 0.03  & std & no  & yes & 0.18 & 17  & 43   &  0.03/0.93 & 0.8 & 0.9\\
P-irr       &      &     &       &       &     &       &     & yes & no  & 0.15 & 16  & 90   &  0.10/0.90 & 1.4 & 2.2\\ 
P-rec-irr   &      &     &       &       &     &       &     & no  & no  & 0.17 & 17  & 47   &  0.03/0.93 & 0.9 & 0.6\\
P-tau       &      &     &       &       &     &       &     & yes & yes & 0.21 & 17  & 76   &  0.13/0.87 & 2.1 & 2.7\\
\hline
\end{tabular}
\end{minipage}
\end{table*}

The implementation of angular momentum evolution in FLASH is described in \citet{FM12}.
Viscous heating is handled as an explicit source term in the energy equation, while the
diffusive momentum flux is added to the advective fluxes obtained during the Riemann solver
step. Tests of this implementation are provided in \citet{FM12}.

\subsection{Initial Conditions}
\label{s:initial_conditions}

The initial condition is an equilibrium torus with constant 
entropy, angular momentum, and electron fraction (e.g., \citealt{PP84,hawley2000}).  
The initial solution is obtained by first finding the local density
given the entropy, electron fraction, and position in the disk,
\begin{eqnarray}
\label{eq:pp_general}
w(\rho,r,\theta)\,\big|_{s,Y_e} & = & 
\frac{GM_{\rm BH}}{R_0}\left[\frac{R_0}{r-R_{\rm S}}\nonumber\right.\\
&& \left.-\frac{1}{2}\frac{R_0^2}{(r\sin\theta)^2}\left(\frac{R_0}{r-R_{\rm S}}\right)^2- \frac{1}{2d}\right],
\end{eqnarray}
where $w = e_{\rm int} + p/\rho$ is the specific enthalpy of the fluid,
and $d$ is the torus distortion parameter. The resulting density
distribution is then integrated to obtain the disk mass. The procedure is
repeated, changing the entropy, until the desired torus mass is obtained.
The distortion parameter $d$ controls the thermal content of the
solution; for fixed mass and electron fraction, it is uniquely determined
by demanding a given entropy or a value of $H/R = c_i/(r\Omega_K)$ at the
point of maximum density.

Evolving this equilibrium torus against a low-density background results in 
the emergence a shock from the edges of the disk due to strong gradients. 
To avoid confusing this initial transient with the viscously-driven outflow 
we are interested in, we evolve the initial torus for 100 orbital
times at $r=R_0$ without angular momentum transport or energy source terms. At this time
a power-law density profile has been established at large radii, and the total mass 
in torus material has decreased by a small amount ($< 1\%$), due to to the spread to larger radii.
The mass accretion rate at the ISCO due to numerical viscosity is $\sim 10^{-5}M_\sun$~s$^{-1}$.
We then reset all positive radial velocities to zero and take this as the initial condition 
for the simulations.

A realistic merger will produce a torus with a more general radial distribution
of mass, angular momentum, and thermal energy, than we have assumed. 
However, we do not expect our results to be overly sensitive to the initial thermal
distribution of the torus, since the temperature profile will equilibrate
on a thermal timescale $t_{\rm cool} \sim t_{\rm visc}(H/R)^{2} \sim 0.1t_{\rm visc}$.

The torus is surrounded by an ambient medium of constant density 
with a value $\sim 20\%$ higher than the floor of density (\S\ref{s:numerical}),
and pressure distribution $p_{\rm amb} = -\Phi\rho_{\rm amb}$ \citep{stone1999}.
The magnitude of the ambient density is chosen as the lowest value that does 
not cause numerical problems in the inner regions of the polar funnel.

\subsection{Models Evolved}
\label{s:models}

We evolve a number of models that explore the effect of (1) varying 
fundamental parameters in the system, and (2) artificially
removing critical processes for outflow generation. All models are shown in 
Table~\ref{t:models}.

The $S$-series of models include all the physics, and vary
parameters around a fiducial model with $M_{\rm BH}=3M_\sun$,
$\alpha = 0.03$, $M_t = 0.03M_\sun$, $Y_e = 0.1$, and $s = 8 $~k$_B$ per baryon, with
standard resolution (\S\ref{s:numerical}).
We call this model S-def, and it is intended to roughly match
the parameters of model B10 PaWi of \citet[ cf. their Figure~14]{ruffert1999}. 
This model is also
marginally optically thick to neutrinos at early times, becoming transparent
(as measured by the minimum optical depth $\tau_{\rm peak}$ [eq.~{\ref{eq:tau_peak_def}}]) after 5
orbits at $r=R_0$.
Models in this series explore the effect on the outflows of varying the torus mass (S-m0.01, S-m0.10), 
torus radius (S-r75), black hole mass (S-M10), initial electron fraction 
(S-y0.05 and S-y0.150), initial entropy (S-s6 and S-s10), magnitude of the viscous 
stress tensor (S-v0.01 and S-0.10), and resolution (S-hr). Models in this series are
evolved for $1000$ orbits at $r=R_0$, except S-def, which is evolved for $3000$ orbits at
the same location.

The $P$-series explores the effect of suppressing
physical processes relative to model S-def, to study the effect on the outflow generation. 
In model P-rec, we remove the binding energy of alpha particles by 
setting $Q_\alpha=0$ in equation~(\ref{eq:recombination_energy}) in
both the initial condition and the subsequent evolution. The series also
explores the effect of shutting down self-irradiation (P-irr), as well as the effect
of suppressing both recombination and self-irradiation simultaneously (P-rec-irr).
We also test the effect of the optical depth corrections to self-irradiation, by
setting $\tau_{\rm i,irr} = 0$ (eq.~[\ref{eq:tau_irr}]) in model P-tau (the global
suppression of neutrino emission by $\tau_{\rm peak}$ [eq.~\ref{eq:tau_peak_def}] 
is still included). All models in this series are evolved for $3000$ orbits at $r=R_0$.

\begin{figure}
\includegraphics*[width=\columnwidth]{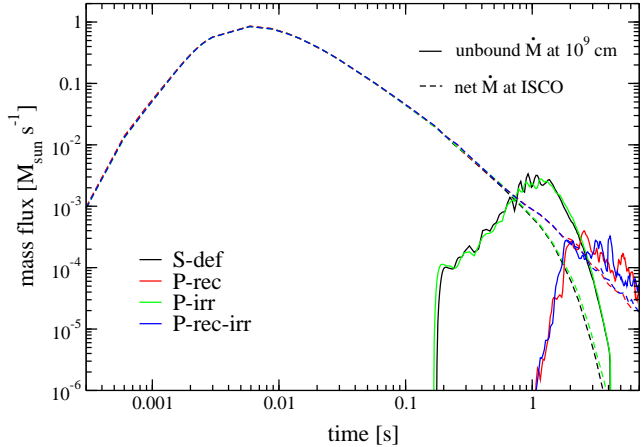}
\caption{Angle-integrated mass outflow rate in unbound material at $r=10^9$~cm (solid lines)
and net mass accretion rate at the ISCO (dashed lines) as a function of time for model S\_def (black).
Also shown are models that suppress the nuclear binding energy of alpha particles
(P-rec, red), self-irradiation (P-irr, green), or both (P-rec-irr, blue).
The orbital time at $r=R_0$ is $2.9$~ms.}
\label{f:mass_flux_evolution}
\end{figure}

\section{Results}
\label{s:results}

\subsection{Outflow generation}

All tori undergo the same basic evolutionary stages.  The disk spreads in radius
due to angular momentum transport, and the internal energy is modified by viscous
heating, nuclear recombination, and neutrino source terms. The accretion rate at the ISCO reaches 
a peak magnitude at time $\sim 0.1/\alpha$ orbits at $r=R_0$, later decaying as a power-law in time 
after $\sim 1/\alpha$ orbits at the same location (Figure~\ref{f:mass_flux_evolution}).

\begin{figure}
\includegraphics*[width=\columnwidth]{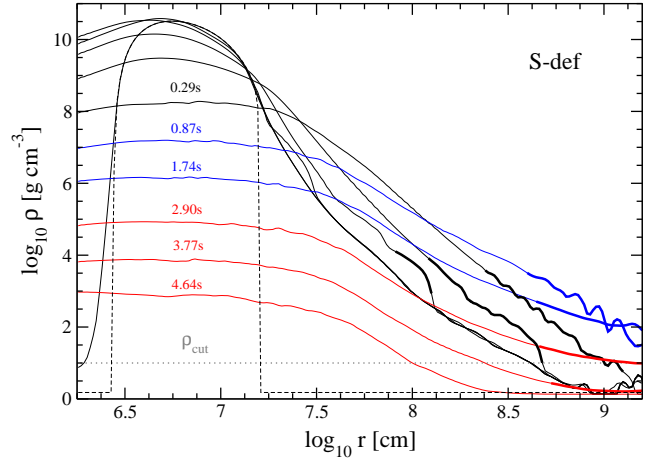}
\caption{Evolution of the angle-averaged density in model S\_def, illustrating the 
generation of the outflow. Curves correspond to times chosen in multiples of the
orbital time at $r=R_0$ ($2.9$~ms), showing $0$, $3$, $10$, $30$, $100$, $300$, $600$, $1000$,
$1300$, $1600$ orbits, as labeled.
The bold curve segments correspond to material that has positive energy (eq.~[\ref{eq:etot_definition}]).
The horizontal dotted line marks the density at which source terms are suppressed 
for numerical reasons, and the dashed line shows the equilibrium torus before relaxation for the
initial condition (\S\ref{s:initial_conditions}).}
\label{f:density_evolution}
\end{figure}

The inner regions of the disk are initially sufficiently hot for neutrino
cooling to balance viscous heating.  In contrast, the outer regions are colder,
and viscous heating dominates over neutrino cooling (e.g., MPQ09).  
This net heating, combined with the transport of angular momentum to larger radii, 
causes an expansion in the outer regions of the disk as shown in 
Figure~\ref{f:density_evolution} for model S-def. The outermost
regions achieve positive specific energy
\begin{equation}
\label{eq:etot_definition}
e_{\rm tot} = \frac{1}{2}\mathbf{v}_p^2 + \frac{1}{2}\frac{\ell_z^2}{(r\sin\theta)^2} + e_{\rm int} + \Phi 
\end{equation}
and are therefore unbound from the central BH.

The mass outflow rate in unbound material at $r=10^9$~cm for model S-def is
shown in Figure~\ref{f:mass_flux_evolution}.  It rises from low values
to a peak at $\sim 1$~s or ($\sim$~400 orbits at $r=R_0$) that exceeds the magnitude
of the mass loss at the ISCO.  This outflow is much denser than that
generated by the initial expansion of the equilibrium torus in the low-density ambient medium,
and which generates the background power-law density profile (Figure~\ref{f:density_evolution}).

The relative contribution of viscous heating, nuclear recombination, and self-irradiation
to the outflow generation is examined in Figure~\ref{f:mass_flux_evolution}. The mass accretion rate
and outflow rate of the default model S-def is compared with those from models that suppress
the nuclear recombination  energy (P-rec), self-irradiation (P-irr), and both nuclear energy and self-irradiation
simultaneously (P-rec-irr). The recombination of alpha particles has the strongest
effect on the disk evolution, by (1) causing a more energetic and massive outflow 
to arise earlier, and (2) by strongly suppressing the mass accretion rate at the ISCO. 
This drop in the accretion rate at late times is consistent with the results of \citet{Lee+09}.

\begin{figure}
\includegraphics*[width=\columnwidth]{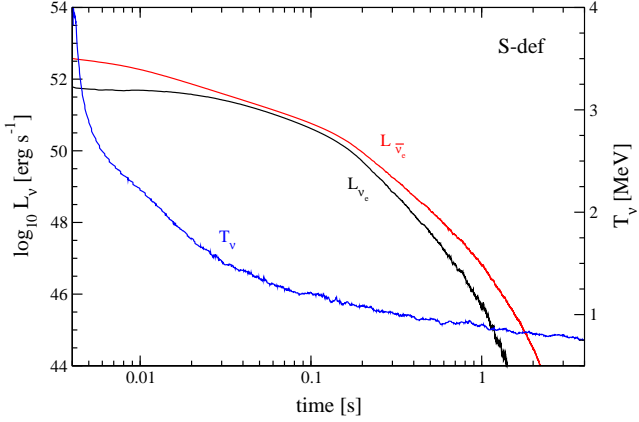}
\caption{Time evolution of the neutrino and antineutrino luminosities (black and red, 
respectively) and neutrino temperature (blue) for model S-def.}
\label{f:neutrinos_time}
\end{figure}

\begin{figure}
\includegraphics*[width=\columnwidth]{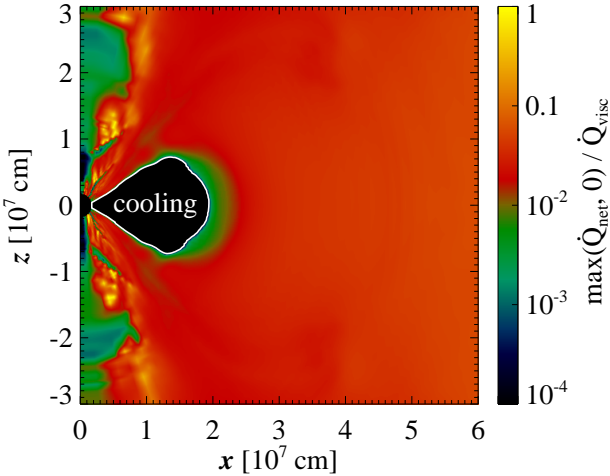}
\caption{Ratio of neutrino heating to viscous heating $\dot{Q}_{\rm visc} = \mathbb{T}:\mathbb{T}/(\nu\rho^2)$ 
(eq.~[\ref{eq:energy_conservation}]) in model S-def-hr at time 0.058s
(orbit 20 at $r=R_0$). The white contour shows the surface $\dot{Q}_{\rm net} = 0$, inside
which neutrino cooling dominates.}
\label{f:heating_ratio}
\end{figure}

Self-irradiation has only a minor effect on the energetics of the system. In model S-def, 
neutrino heating is always smaller than viscous heating by at least a factor of several.
This sub-dominance of self-irradiation can be traced to a few causes: (1) the neutrino
luminosities are high only for a limited period of time, because both the mass
and the temperature of the torus decrease with time, as shown in 
Figure~\ref{f:neutrinos_time} for model S-def; (2) self-irradiation is the
strongest at the time where the torus becomes optically thin, and affects mostly
vertical, low-density regions of the inner disk; and (3) the temperature
in the emission region decreases with time (Fig.~\ref{f:neutrinos_time}), 
resulting in lower mean energies of the neutrinos and thus 
a decreasing energy deposition rate. For illustration, a snapshot of the ratio of neutrino heating to 
viscous heating in model S-def-hr is shown in Figure~\ref{f:heating_ratio}. The effect is
not entirely negligible however, as turning off optical depth corrections leads to
$\sim 20\%$ changes in the electron fraction and the asymptotic velocity (\S\ref{s:outflows}).

The total mass accreted through the ISCO radius and the mass with $e_{\rm
tot}>0$ ejected beyond a radius of $10^9$~cm are given in Table~\ref{t:results}
as fractions of the initial torus mass. In most models, approximately 10 per
cent of the torus mass is ejected, with the rest accreting onto the BH.  
One exception is the model S-r75 with a larger initial torus radius, which ejects
twice as much mass as the default model. Another exception is model S-M10, which
while having a large disk radius, also has a higher BH mass, which leads to
smaller mass ejection by $\sim 50\%$.
The other two outliers are the
models that suppress nuclear recombination (P-rec and P-rec-irr), for which
$\sim 3$ per cent of the mass is ejected (c.f.
Figure~\ref{f:mass_flux_evolution}). Doubling the resolution in each coordinate
direction leads to insignificant quantitative differences, so we consider our
results converged.
Given the range of disk masses explored,
the ejecta mass from delayed outflows can thus range from $10^{-3}$ to
$10^{-2}M_{\odot}$. 

\begin{figure*}
\includegraphics*[width=\textwidth]{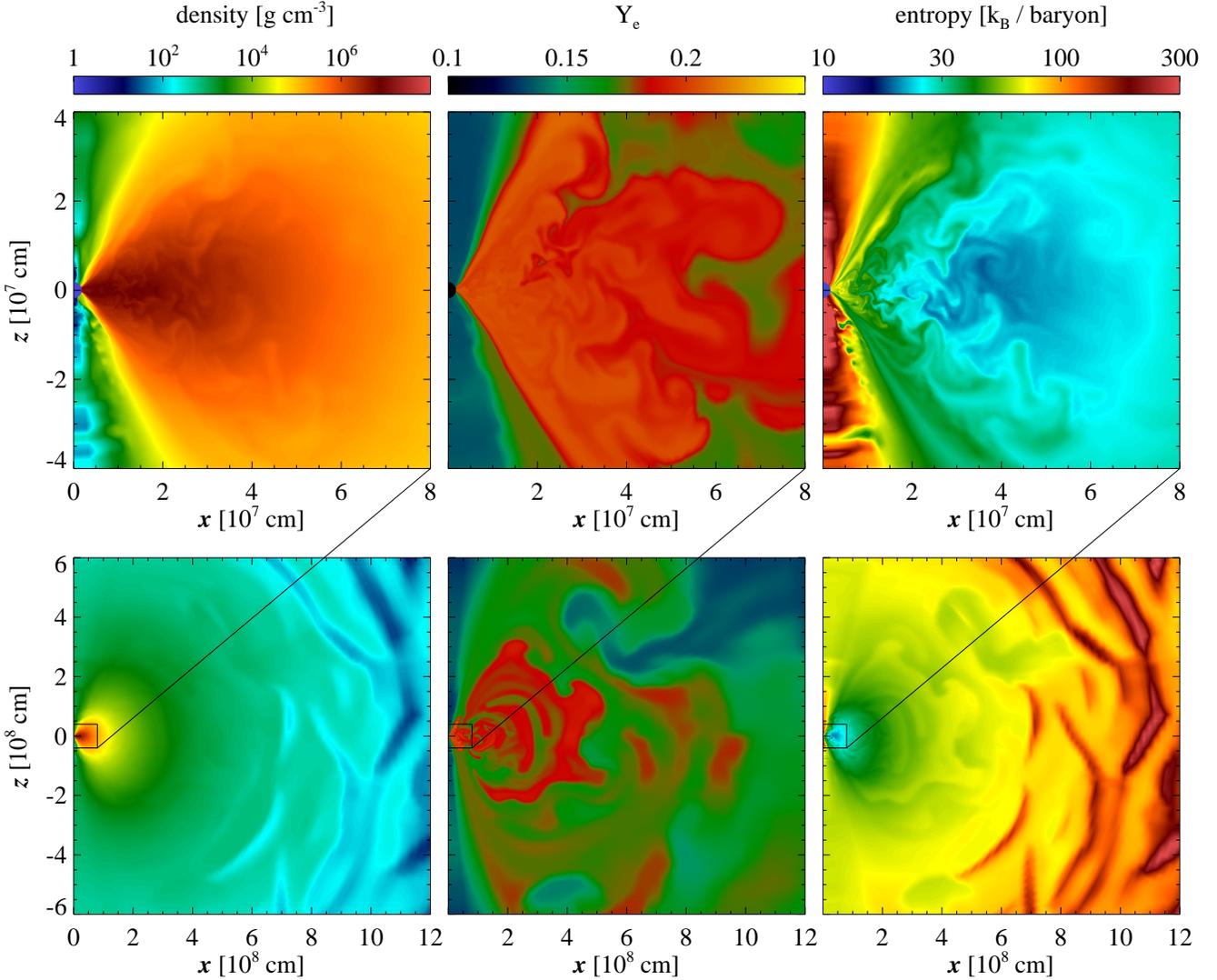}
\caption{Color maps of density (left), electron fraction (middle), and entropy (right)
in model S-def-hr at time 1.16s (orbit 400 at $r=R_0$), illustrating the wind morphology. 
Upper panels show zoomed in regions of the panels directly below, as indicated by the boxes.
An animated version of this figure is available in the online journal.}
\label{f:mosaic}
\end{figure*}

The morphology of the ejecta is quasi-spherical, with most of the material being
ejected within $\sim 60^\circ$ of the disk midplane, as shown in Figure~\ref{f:mosaic}.
The bulk of the disk has moderate entropy ($\sim 20$), particularly in regions
relevant for nucleosynthesis (\S\ref{s:outflows}).
At large radii ($\sim 10^9$~cm), the outflowing material develops shocks that raise its entropy.
Large scale instabilities are also seeded by convection in the disk, which cannot cool
efficiently once its accretion rate falls below $\dot{M}_{\rm ign}$ (eq.~[\ref{eq:Mdotign}]).
The funnel within a few tens of degrees from the polar axis has different properties; we do 
not study them here given our very approximate treatment of general relativity, and our
omission of energy deposition by neutrino pair annihilation.

\subsection{Outflow properties}
\label{s:outflows}

\begin{figure}
\includegraphics*[width=\columnwidth]{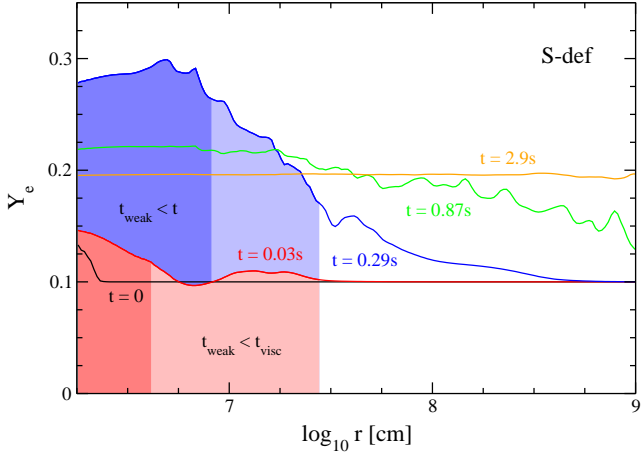}
\caption{Angle-averaged, mass-weighted electron fraction at selected times for model S-def, as labeled. 
The dark-filled region corresponds
to the weak equilibration time $t_{\rm weak} = Y_e/|\Gamma_{\rm net}|$ being smaller than the curve time, 
while the light-filled region is such that $t_{\rm weak}$ is shorter than the viscous time $t_{\rm visc}$.}
\label{f:ye_freezout}
\end{figure}

At early times and in the innermost regions of the disk ($r\lesssim 100$ km),
weak interactions are faster than the viscous time, and the electron fraction
reaches its equilibrium value given the local density and temperature
(e.g.,~\citealt{Beloborodov03}).  As the disk evolves, the temperature and
density decrease to the point that weak interactions operate slower than the
viscous time and the electron fraction freezes out, becoming an advected
quantity only (MPQ09). This process is illustrated for our fiducial model in
Figure~\ref{f:ye_freezout}.

In the inner regions of the disk, the equilibrium electron fraction is
initially small because electrons are more abundant than positrons when the
disk is degenerate.
However, with time $Y_e$ rises to
a maximum value $\sim 0.3$ because the disk becomes less degenerate as the
density decreases. Most of this high-$Y_e$ material is accreted onto the BH,
however.  The outer parts of the disk that eventually become unbound never
achieve such high electron fractions and are instead ejected with $Y_e \lesssim
0.2$.

\begin{figure}
\includegraphics*[width=\columnwidth]{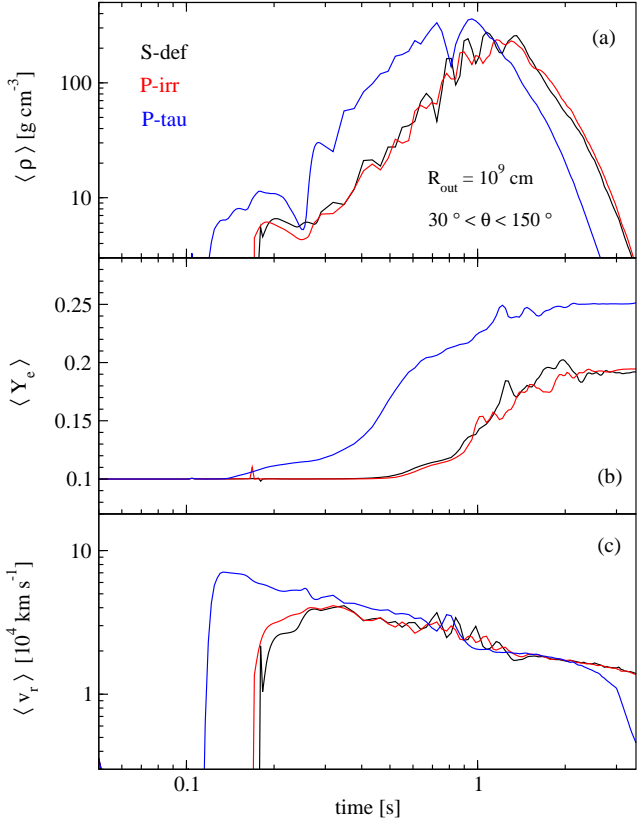}
\caption{Instantaneous mass-flux-weighted quantities at a radius $R_{\rm out}=10^9$~cm 
for models S-def, P-irr (no self-irradiation), and P-tau (no optical depth suppression). 
Shown are the density $\rho$ (a), electron fraction $Y_e$ (b), and 
velocity $v_r$ (c). The average is conducted over polar angles within $60^\circ$ of the equator.}
\label{f:fluxes_outflow}
\end{figure}

At a given radius, the wind material is not homogeneous in composition, but instead
shows some variation around mean values.
Figure~\ref{f:fluxes_outflow} shows mass-flux weighted quantities crossing a
spherical surface at the fixed radius $R_{\rm out}=10^9$~cm in the models S-def and
P-irr.  Quantities are computed according to
\begin{equation}
\langle A(R_{\rm out},t) \rangle = 
\frac{\int_{\theta_{\rm min}}^{\theta_{\rm max}}\sin\theta\totd\theta\,R_{\rm out}^2\rho v_r A(R_{\rm out},\theta,t)}{
				\int_{\theta_{\rm min}}^{\theta_{\rm max}}\sin\theta\totd\theta\,R_{\rm out}^2\rho v_r},
\end{equation}
where $A(R_{\rm out},\theta\,, t)$ is a generic scalar quantity, and
$[\theta_{\rm min},\theta_{\rm max}]$ is the angular range of the meridional
integral, chosen as a wedge within $60^\circ$ of the midplane (c.f. Figure~\ref{f:mosaic}).  
The bulk of the ejected material has $Y_e\sim 0.2$ and velocities $v_r \sim 20,000$~km~s$^{-1}$. 

Heavy nuclei generation occurs in regions where 
$T\sim 5\times 10^9$~K \citep{Hoffman+97}. In the fiducial model S-def, this
corresponds to a range of radii $200-400$~km on the midplane depending on the
phase in the disk evolution.
To quantify the thermodynamic properties relevant for nucleosynthesis
in the disk, we compute time- and angle-integrated mass-weighted mean values
\begin{equation}
\label{eq:mean_value}
\bar{A}(R_{\rm nuc}) = \frac{\int \totd t\int \sin\theta\totd\theta\, R_{\rm nuc}^2\rho v_r\, A(R_{\rm nuc},\theta,t)}{
                \int \totd t\int\sin\theta\totd \theta\, R_{\rm nuc}^2\rho v_r},
\end{equation}
where the radius $R_{\rm nuc}$ is chosen so that $\bar{T} \simeq 5\times 10^9$~K.
The resulting mean values of the electron fraction, entropy, and expansion time
at $R_{\rm nuc}$ (within $60^\circ$ of the midplane) are shown for all
models in Table~\ref{t:results}.

\begin{figure*}
\includegraphics*[width=\textwidth]{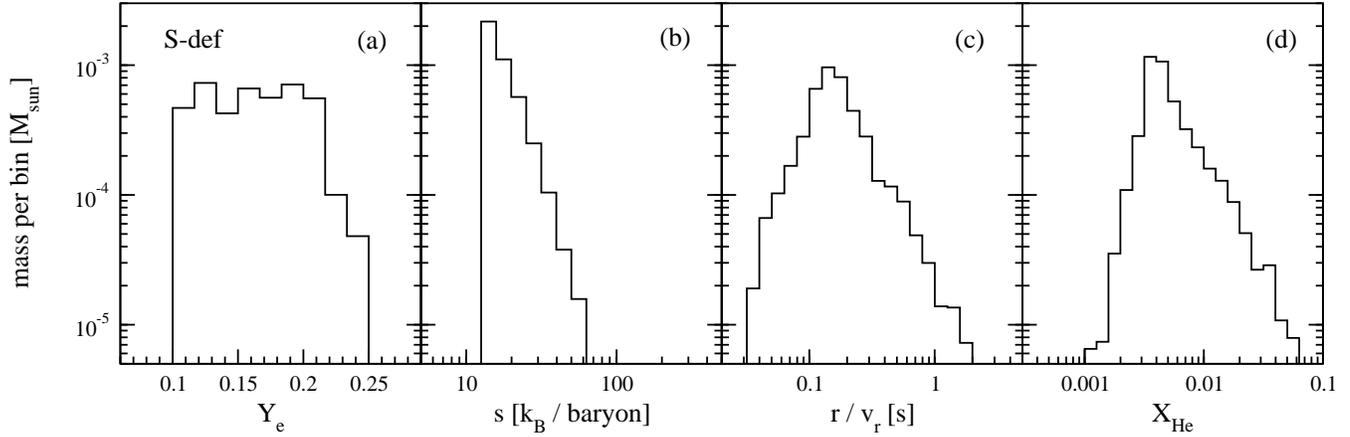}
\caption{Thermodynamic properties of disk material for model S-def at a radius $R_{\rm nuc}=400$~km, where
$\bar{T}\simeq 5\times 10^9$~K. The histograms are constructed by considering all material that
crosses this radius, within $60^\circ$ of the midplane, over the entire disk evolution. Shown are the 
electron fraction $Y_e$ (a), entropy $S$ (b), expansion time $t_{\rm exp} = r/v_r$ (c), and
final helium mass fraction (d) computed from equation~(\ref{eq:XHe}).}
\label{f:histogram_outflow}
\end{figure*}

The composition is largely insensitive to variations in the model parameters,
with typical mean values $\bar{Y}_e \sim 0.2$, $\bar{S}\sim 20k_{\rm B}$
per baryon, and expansion timescales $t_{\rm exp} = R_{\rm nuc}/\bar{v}_r \sim 0.1$~s. 
The magnitude of the spread from the mean value is illustrated in Figure~\ref{f:histogram_outflow},
which breaks up the mean values of $Y_e$, entropy, and expansion time for model S-def.

The value of $\bar{Y}_e$ is relatively insensitive to the initial electron fraction
because most of the disk mass achieves weak equilibrium,
erasing its initial composition.  The largest difference
in $\bar{Y}_e$ occurs due to changes in the initial entropy of the disk and the
magnitude of the viscosity.  The entropy of the torus is related to the level
of degeneracy, which in turn determines the value of the equilibrium electron
fraction (e.g., \citealt{Beloborodov03}). The magnitude of the viscosity
determines the ratio of the weak equilibration time $t_{\rm weak}$ to the
viscous times $t_{\rm visc}$.  A smaller viscosity implies that these two
timescales become equal at a large radius, which increases the amount of
processed $Y_e$ that enters the outflow.  This modest dependence of $Y_e$ on
$\alpha$ is, however, mitigated by two facts: (1) the disk is regulated by neutrino cooling to moderate degeneracy (CB07), independent of the magnitude of viscous heating; (2) the disk becomes radiatively inefficient (and hence thicker and less degenerate) on the same timescale that $Y_e$ freezes out.  This coincidence results from the fact that $e^{-}/e^{+}$
captures both cool the disk and change $Y_e$ \citep{Metzger+09a}.

The effects of self-irradiation on the ejecta composition are illustrated by
Figure~\ref{f:fluxes_outflow}, which compares the polar mass-flux
weighted composition for model S-def, P-irr (which suppresses self-irradiation),
and P-tau (which turns off optical depth corrections to self-irradiation).
Including self-irradiation with optical depth correction leads to a nearly
identical outcome than suppressing irradiation altogether. 
As with the small energetic relevance of neutrino heating,
changes in the composition due to neutrino irradiation are important only
around the time when the disk becomes transparent to neutrinos and its
luminosity is still high.  This timescale of peak irradiation is somewhat
shorter than that required to significantly change the composition by weak
interactions. The largest possible effect of neutrino heating on the outflow
is probed by model P-tau, which completely suppresses optical depth corrections
to the self-irradiation luminosity (while still globally suppressing the cooling rates).
The outflow begins earlier, achieves asymptotic velocities that are $\sim 20\%$ larger, 
and has a mean $Y_e$ that is 0.05 larger than the default model. Such moderate
changes given an unrealistically large neutrino absorption rate reaffirms our
conclusion that self-irradiation is only a small correction to the overall dynamics
for low mass disks.

The expansion velocity at larger radii is remarkably constant, sensitive only to the inclusion of the 
nuclear binding energy of alpha particles.
Note however that these values likely underestimate the true expansion velocities 
by a factor $\sim 1.4-1.7$ since we do not include the additional energy of $\sim $ few 
MeV nucleon$^{-1}$ released in forming seed particles and (on somewhat longer timescales 
$\sim$ 1 s) heavy $r$-process nuclei.

\section{Discussion}
\label{s:discussion}

\subsection{Nucleosynthesis in Disk Outflows}
\label{s:nucleosynthesis}

Appendix \ref{s:nucleo} reviews the conditions for heavy element
nucleosynthesis in hot outflows, such as those from NS merger accretion disks,
and provides analytic formulae for the fraction of the ejected mass synthesized
into heavy $r$-process nuclei $X_{\rm h}$ and the fraction $X_{\alpha} = 1 -
X_{\rm h}$ that remains in $\alpha$-particles (eq.~[\ref{eq:XHe}]).  Given the
low electron fraction $Y_e \lesssim 0.2$ of the disk outflows found in our
calculations, we conclude that most of the ejecta will go into 2nd and 3rd peak
$r$-process elements with mass number $A \sim 130-200$ ($X_{\rm h} \simeq 1$).
A much smaller fraction, X$_{\rm He} \sim 10^{-2}$, will remain 
in $^{4}$He, which may have implications for the
spectroscopic signatures of these events ($\S\ref{s:kilonova}$).  

The average Galactic production rate of $r$-process nuclei due to accretion
disk outflows from binary NS mergers is given by 
\begin{eqnarray}
\dot{M}_{\rm r} &=& \mathcal{R}_{\rm NS^{2}}f_{\rm ej}\bar{M}_{t} \nonumber \\
&=& 10^{-7}M_{\odot}{\rm yr^{-1}}\left(\frac{\mathcal{R}_{\rm
NS^{2}}}{10^{-4}{\rm yr^{-1}}}\right)\left(\frac{f_{\rm
ej}}{0.1}\right)\left(\frac{\bar{M}_{\rm t}}{10^{-2}M_{\odot}}\right),
\nonumber \\ \label{eq:Mdotr} 
\end{eqnarray} 
where $\mathcal{R}_{\rm NS^{2}}$
is the rate of NS-NS mergers scaled to its estimated value in the Milky Way
(uncertain by at least an order of magnitude; \citealt{Kalogera+04}),
$\bar{M}_{\rm t}$ is the average mass of the initial accretion torus, and
$f_{\rm ej} \sim 0.1$ is the fraction of $M_{\rm t}$ ejected in outflows,
scaled to the characteristic value derived from our numerical simulations.
Equation (\ref{eq:Mdotr}) should be compared with the `observed' rate
$\dot{M}_{\rm r}^{\rm obs} \sim 5\times 10^{-7}M_{\odot}$ yr$^{-1}$ required to
explain the abundances of heavy $A \gtrsim 130$ $r$-process elements produced
over the age of the Galaxy  (\citealt{Qian00}).  This shows that NS merger disk
outflows consistitute a potentially significant $r$-process source.

Although the idea that NS mergers are a promising $r$-process source is not new
(\citealt{Lattimer&Schramm74}), previous studies have focused almost
exclusively on nucleosynthesis of the dynamical ejecta.  

Our simulations show no clear evidence for outflows powered primarily by neutrino
heating. Such a wind was previously anticipated to dominate mass loss from small
radii in the disk (e.g.~\citealt{Metzger+08b}; \citealt{Surman+08}; \citealt{Wanajo&Janka12}).  Because
neutrinos and antineutrinos from the disk have similar luminosities and mean
energies, neutrino absorption is expected to drive $Y_e$ to a value $\sim 0.5$.
Although our simulations do show unbound polar outflows, the electron fraction
of this material is increased only slightly by neutrino irradiation
(Fig.~\ref{f:fluxes_outflow}), since viscous heating dominates
the unbinding of matter from the disk (Fig.~\ref{f:heating_ratio}).  This hierarchy
is contingent upon our assumption of an $\alpha-$viscosity, which may not
reflect the true vertical distribution of turbulent dissipation in the disk.
Our approximate treatment of neutrino physics prevents us from definitively
ruling out a strong role of self-irradiation.
However, if an even larger fraction of the dissipation occurs in the
disk corona in more realistic magnetized disks (e.g.~\citealt{Hirose+06}), then
the dominance of `viscous'-driven (low $Y_e$) polar outflows may turn out to 
be a robust feature of the launching mechanism.  Regardless of the composition of the [possibly neutrino-driven] winds from small radii in the disk, the late outflows powered by $\alpha$ recombination and viscous heating almost certainly dominate the total (time-integrated) mass loss.  

\subsection{Radioactively-Powered Emission}
\label{s:kilonova}

Radioactive decay of $r$-process elements synthesized in the disk outflows
gives rise to an electromagnetic transient similar to a dim supernova
(e.g.~\citealt{Li&Paczynski98}).  Although the disk outflows are mildly
anisotropic (Fig.~\ref{f:mosaic}), the ejecta will become increasingly spherical as it
expands homologously with a characteristic velocity 
$\bar{v}_r \simeq 0.1c$ (Table~\ref{t:models}). 
 Most of the energy released by the $r$-process occurs
on a timescale $\sim$ seconds, but this heating is lost to adiabatic expansion since
the outflow is highly optically thick at this early stage.  Nevertheless,
radioactive decay continues to add energy to the ejecta, at an approximately
constant rate per logarithmic time (e.g.~\citealt{Metzger+10}).
Electromagnetic emission peaks only once the timescale for photons to diffuse
through the ejecta $t_{\rm d} \propto \kappa M_{\rm ej}/cr$ becomes shorter
than the expansion timescale $t_{\rm exp} = r/\bar{v}_r$.  This occurs on a
timescale (\citealt{Metzger+10}, eqs.~[3],[4])
\begin{equation}
t_{\rm p} \approx 7{\rm d}\left(\frac{\bar{v}_r}{0.1c}\right)^{1/2}\left(\frac{M_{\rm ej}}{10^{-2}M_{\odot}}\right)^{1/2}
\left(\frac{\kappa}{10{\rm \,cm^{2}g^{-1}}}\right)^{1/2}
\end{equation}
resulting in emission with a characteristic peak luminosity
\begin{eqnarray}
L_{\rm p} &\simeq& 2.4\times 10^{40}{\rm erg\,s^{-1}}\left(\frac{\bar{v}_r}{0.1c}\right)^{-1/2} \nonumber \\
&& \qquad\times\left(\frac{M_{\rm ej}}{10^{-2}M_{\odot}}\right)^{1/2}\left(\frac{\kappa}{10{\rm \,cm^{2}g^{-1}}}\right)^{-1/2},
\end{eqnarray}
where $\kappa$ is the opacity, scaled to a value characteristic of the line
opacities of Lanthanide elements, given their expected abundance in the
$r$-process ejecta \citep{Kasen+13}.

Because the composition of the disk outflows (predominantly heavy $r$-process
nuclei) is similar to that of the dynamically ejected tidal tails, it may be
observationally challenging to distinguish their contribution to the transient
emission.  One unique feature of the disk outflows, however, is the presence of
a modest amount of helium ($X_{\rm He} \sim 10^{-2}$), which is not present
in the dynamical ejecta due to its much lower entropy $S \lesssim$ $k_b$
baryon$^{-1}$ (since $X_{\rm He} \propto S^{3/2}$; eq.~[\ref{eq:XHe}]).  

Even a small quantity of He may be detectable if emission/absorption lines (e.g.~He I $1.083\mu$m and He I $2.058\mu$m) are produced as
the result non-thermal excitation, induced by energetic $\gamma$-rays,
$\beta$-decay electrons and fission products from decaying $r$-process
elements.  This is in analogy to the mechanism responsible for producing He
lines in core-collapse SNe as the result of mixing radioactive $^{56}$Ni into
the outer layers of the ejecta (e.g.~\citealt{Lucy91}, \citealt{Dessart+12}).
Though a potentially promising diagnostic of conditions close to the merger
site, future work is required to quantify the strength of He emission lines in
kilonovae.

\subsection{Effects of a Long-Lived Hypermassive NS}
\label{s:HMNS}

In the case of NS-NS mergers, our calculations implicitly assume that a BH
forms within a few dynamical times after coalescence.  If, however,
the collapse is significantly delayed due to, e.g., thermal pressure and/or
differential rotational support (\citealt{Morrison+04, OConnor&Ott11,
Paschalidis+12, Lehner+12}), then the effects of neutrino irradiation from the
hyper-massive NS (HMNS) could qualitatively alter the properties of the disk
evolution and its outflows.

To significantly alter the disk {\it dynamics}, neutrino emission from the HMNS
would need to maintain both a high flux level, and a high mean neutrino energy,
for a timescale longer than both (1) the time required to achieve neutrino
transparency; and (2) the thermal time of the disk, such that the internal
energy is raised sufficiently to unbind the disk.  Such neutrino-driven
outflows necessarily occur via thermal evaporation
(e.g.~\citealt{McLaughlin&Surman05}; \citealt{Metzger+08b};
\citealt{Surman+08}; \citealt{Dessart+09}), as in the case of proto-neutron
star winds, since momentum transfer from neutrinos is negligible.\footnote{Even
a neutrino luminosity as large as $L_{\nu} \sim 10^{53}$~erg~s$^{-1}$ is
sub-Eddington.}

Even if neutrino irradiation from the HMNS is insufficient to alter the
dynamics of the disk, it may still alter its composition, especially in outer
parts of the disk where the rate of electron/positron captures is slow.  To the
extent that the HMNSs formed in binary NS merger are similar to the
proto-neutron stars formed in core collapse supernovae, the neutrino and
antineutrino luminosities and temperatures are expected to be similar.  The net
effect of HMNS irradiation on the disk would thus be to raise $Y_e$ in both the
disk midplane and in subsequent outflows (\citealt{Dessart+09}; see
\citealt{Metzger+09b} and \citealt{ Darbha+10} for an analogous scenario
following the accretion-induced collapse of a white dwarf).  Spectroscopic or
photometric signatures of elements synthesized exclusively in proton-rich
ejecta of kilonovae (e.g.,~Fe or $^{56}$Ni) could thus be taken as evidence for
a long-lived HMNS.  One significant difference between an $r$-process-rich and
an Fe-rich outflow is the opacity of the ejecta, which drastically changes the
rise time and peak luminosity of the light curve (\citealt{Barnes&Kasen13,tanaka2013}).
Future work is necessary to quantify how long such a HMNS must survive in order
to appreciably alter the outflow properties from the purely BH accretion
scenario studied in this paper.

\subsection{Comparison with Previous Work}

\citet{ruffert1999} first studied the evolution of disks formed after 
NS coalescence in 3D using a realistic equation of state, a finite volume hydrodynamic
method, Newtonian and pseudo-Newtonian potentials, and a neutrino leakage scheme \citep{ruffert1996}. 
Follow up work by \citet{setiawan2004} and \citet{setiawan2006} explored the 
effect of including an $\alpha$ viscosity over a wide range of disk masses, 
finding that neutrino emission is 
sensitive to the magnitude of $\alpha$. Our results cannot be directly compared to theirs, however,
because their models were evolved only up to $70$~ms. Also, due to the equation of state used,  
their density floor was set at $10^8$~g~cm$^{-3}$, which would make it hard to
develop any viscously driven outflow (c.f. Figure~\ref{f:density_evolution}).

The interplay between magnetohydrodynamic and general relativistic effects was
explored by \citet{shibata2007}, finding that if poloidal fields and differential
rotation are present, the magnetorotational instability drives turbulence that 
transports angular momentum and heats the disk. Due to the MHD algorithm
employed, these models also require a floor of density $\sim 10^6$~g~cm$^{-3}$,
which again prevents the development of viscously driven winds out to large radius.
The effect of gray radiation transport has been investigated by \citet{shibata2012}, and
general relativistic simulations of tori formed in BH-NS mergers have recently
been reported by \citet{deaton2013}.

\citet{lee2005b} evolved tori over timescales
up to a few $100$~ms, using initial conditions from
a merger simulation. They used a smoothed particle hydrodynamics method, and 
included a realistic equation of state, neutrino
optical depth corrections, and the effect of nuclear recombination in the 
energetics of the disk. In contrast to our models, they assumed that the electron fraction
satisfies beta equilibrium, parameterized the NSE abundances by the analytic
relation of \citet{Qian&Woosley96}, included all components of the viscous stress
tensor, and focused on relatively massive tori. \citet{Lee+09}  
extended the evolution to a few seconds, considered a pseudo-Newtonian
potential, and explored a wide range of tori masses and viscosity parameters.
They found that the accretion rate drops sharply at late times as a consequence of including
the recombination energy of alpha particles, an effect that we reproduce
in our models (c.f. Figure~\ref{f:mass_flux_evolution}).

The properties of the disk evolution and outflows found here are in
qualitative agreement with the 1D model of MPQ09.  MPQ09 predicted a somewhat
larger mass of unbound material, and a distribution of electron fraction
extending to higher values $Y_e \gtrsim 0.3$.  In hindsight,
this discrepancy can be understood by the fact that MPQ09 counted all
mass in the disk outside the point of $Y_e$ freeze-out as eventually becoming
unbound, whereas our calculations show that only the outer edge of the disk
(where $Y_e$ is lowest) achieves sufficiently high energy to become unbound.  The
temperature and density of the disk midplane in our simulations are also
somewhat lower than in the 1D $\alpha$-disk models of MPQ09 (they find an
inner disk temperature of $\sim 10$ MeV), resulting in less effective neutrino
cooling.  This discrepancy can partially be understood by the fact that
$\alpha$-disk models assume that viscous heating exactly balances cooling,
which is not everywhere achieved in our models, leading the 1D models to
overestimate the midplane density.  Since radiation pressure is moderately
important, lower density implies a lower temperature in vertical hydrostatic
balance.

\section{Conclusions}
\label{s:conclusions}

We have explored the properties of outflows generated on the viscous
time by accretion disks formed in NS-NS and NS-BH mergers, with
an eye on detectable electromagnetic counterparts of the gravitational
wave emission and $r$-process nucleosynthesis. 
Two-dimensional, time-dependent hydrodynamic simulations
with a realistic equation of state, weak interactions, 
self-irradiation in moderately optically-thick environments, and
angular momentum transport via an $\alpha$-viscosity are used to characterize the
driving, composition, and parameter dependencies of these outflows.
Our main findings are as follows:
\newline 

\noindent
1. -- Disks eject $\sim 10\%$ of their mass in an unbound outflow.
      The composition is neutron rich, with typical electron fractions
      $Y_e\sim 0.2$. This value arises because the ejected
      material comes from the outer part of the disk, which is never
      processed to a high equilibrium electron fraction by weak interactions.
      The high-$Y_e$ material in the inner disk regions is accreted onto the BH
      (Fig.~\ref{f:ye_freezout}).  Neutron-rich freeze-out is robust because (a)
      neutrino cooling regulates the disk to be mildly degenerate, which in turn
      regulates the equilibrium $Y_e$ prior to the freeze-out; (b) outflows begin
      once the disk becomes advective and $\alpha-$particles form, both which occur
      nearly simultaneous with freeze-out.  
      \newline

2. -- The outflow composition is relatively robust relative to initial disk
      parameters (Table~\ref{t:models}). In regions where heavy elements 
      form ($T\sim 5\times 10^9$~K), we obtain characteristic 
      entropies $\sim 20\,k_b$ baryon$^{-1}$ and expansion times $\sim 0.1$~s.
      The composition is not homogeneous, with spatial
      as well as temporal variations around the mean value 
      (Figs.~\ref{f:mosaic}, \ref{f:histogram_outflow}).
      The expected nucleosynthetic output consists in 2nd/3rd peak
      $r$-process elements ($A \gtrsim 130$), with a small trace of He ($\sim 1\%$ by
      mass, but $\gtrsim 10\%$ by number), which may generate a spectroscopic signature. 
      \newline

3. -- Small variations $\sim 10\%$ in the electron fraction are
      introduced by changes in the initial torus entropy and
      the magnitude of the viscosity. The entropy of the outflow
      is moderately sensitive to the initial torus mass and
      its initial entropy.
      The amount of mass ejected is sensitive only to the
      radius at which the torus mass resides initially.
      The outflow velocity is remarkably robust, with a characteristic
      value of $\sim 20,000$~km~s$^{-1}$ at $10^9$~cm.
      \newline

4. -- In addition to viscous heating, nuclear recombination is a 
      fundamental component in the generation of the 
      outflow (Fig.~\ref{f:mass_flux_evolution}). Its exclusion results
      in much lower ejected masses and asymptotic velocities. We also
      witness the drop in the accretion rate at late times first seen
      by \citet{Lee+09}.
      \newline

5. -- Self-irradiation appears to be both energetically and compositionally
      sub-dominant. This stems in part from the fact that irradiation is
      maximal around the time when the disk becomes transparent to
      neutrinos, which for most of our low-mass disks corresponds to early
      times in the disk evolution. The subsequent drop in disk temperatures
      and densities lowers both the amount of neutrinos emitted and their
      mean energies (Fig.~[\ref{f:neutrinos_time}]), 
      diminishing the relative importance of neutrino energy 
      deposition as time passes.
      The effects of self-irradiation become noticeable only when completely
      removing optical depth corrections (Fig.~\ref{f:fluxes_outflow}). Even then,
      changes in global quantities are only of the order of $10\%$.
      \newline

There are many possible improvements to the model. Self-consistent
angular momentum transport by magnetohydrodynamic turbulence, magnetocentrifugal
winds, and/or self-gravity (for more massive disks) would yield a first-principles
estimate on the spatial distribution of angular momentum and heating in the disk.
A full general-relativistic code would help address whether the properties of the
inner disk are important, particularly regarding neutrino self-irradiation with
rapidly spinning black holes.
More realistic initial conditions would shed light on whether the initial
angular momentum distribution is of consequence for the timing and asymptotic
properties of the ejecta.  The possible effects of late `fall-back' accretion of tidal tail material (e.g.~\citealt{Metzger+10b}) on the disk evolution and outflows, should also be explored.  More accurate neutrino physics and three-dimensional
modeling would be needed if precise predictions for the outflow properties are necessary.

A more detailed analysis of the nucleosynthesis products and electromagnetic
signal generated by these outflows will be carried out in subsequent work.

\section*{Acknowledgments}

We thank Stephan Rosswog, Thomas Janka, Almudena Arcones, Gabriel Mart\'inez, 
Will East, Andrei Beloborodov, Tony Piro, Eliot Quatert, Dan Kasen, and 
Luke Roberts for stimulating discussions and/or comments on the manuscript.
We also thank the anonymous referee for helpful comments that improved the manuscript.
RF is supported by NSF grant number AST-0807444. BDM acknowledges support from Columbia University.
The software used in this work was in part developed by the DOE NNSA-ASC OASCR Flash Center at the
University of Chicago. Computations were performed at the IAS \emph{Aurora} cluster.

\appendix

\section{Accretion Rate for Opaqueness to Neutrinos}
\label{s:mdot_opaque}

Here we provide an analytic derivation of the critical accretion rate
$\dot{M}_{\rm opaque}$ above which the inner disk becomes opaque to neutrinos.
A steady-state disk is optically thick at radii interior to the point at which the
vertical optical depth to neutrino absorption obeys
\be
\tau_{\nu} \equiv \Sigma\kappa_{\nu}/2 > 1,
\label{eq:taunu_cond}
\ee
where $\Sigma$ is the disk surface density and $\kappa_{\nu}$ is the mean
opacity.   For an accretion disk in steady-state, the accretion rate and
surface density are related by $\dot{M} = 3\pi\nu \Sigma$ at radii much greater
than the inner boundary, where $\nu = \alpha c_{s}H$ is the kinematic
viscosity; $c_{s}$ is the sound speed; $H = c_{s}/\Omega_{\rm K}$ is the disk
scaleheight; and $\Omega_{\rm K} = (GM_{\rm BH}/r^{3})^{1/2}$ is the Keplerian orbital
frequency.  At accretion rates $\dot{M} \gtrsim \dot{M}_{\rm ign}$, the inner
disk is neutrino-cooled, such that the midplane is sufficiently dense
that gas pressure exceeds radiation pressure, and the opacity is
dominated by the absorption of neutrinos on free nuclei (CB07).  In this case
$c_{s} \simeq (4k_{B}T/3m_{p})^{1/2}$ and $\kappa_{\nu} \simeq \kappa_{0}T^{2}$
(where $\kappa_0 \simeq 4.5\times 10^{-39}$ K$^{-2}$cm$^{2}$g$^{-1}$; e.g.~\citealt{DiMatteo+02}).  
Equation (\ref{eq:taunu_cond}) can thus be expressed as a
condition on the accretion rate
\be
\dot{M} > \frac{8\pi\alpha k_{B}}{\kappa_0 m_p \Omega_{K} T}
\label{eq:Mdotcond}
\ee
Now, for an optically thin disk, the midplane temperature is determined by the
balance between viscous heating $\dot{q}_{\rm visc} = (9/4)\nu\Omega_{K}^{2}$
and neutrino cooling $\dot{q}_{\nu} \simeq \dot{q}_{\nu,0}T^{6}$ (due to
e$^{-}$/e$^{+}$ captures on free nuclei, where $\dot{q}_{\nu,0} \simeq 8\times
10^{-43}$T$^{6}$ erg s$^{-1}$g$^{-1}$K$^{-6}$), viz.~
\begin{eqnarray}
T & \simeq & 1.25(\alpha k_{B}/m_p)^{1/5}\Omega_{\rm K}^{1/5}\dot{q}_{\nu,0}^{-1/5}\\
  & \approx & 5.9\times 10^{10}{\rm\,K}\left(\frac{\alpha}{0.03}\right)^{1/5}\left(\frac{M_{\rm BH}}{3M_{\odot}}\right)^{-1/5}\left(\frac{r}{R_{g}}\right)^{-3/10},\nonumber\\
\label{eq:Tprofile}
\end{eqnarray}
where we have now expressed radius in terms of gravitational radii $R_{\rm g} =
GM_{\rm BH}/c^{2}$.  Substituting equation (\ref{eq:Tprofile}) into equation
(\ref{eq:Mdotcond}), we find
\be
\dot{M} > 2\times 10^{-3}M_{\odot}{\rm\,s^{-1}}\,\left(\frac{\alpha}{0.03}\right)^{4/5}\left(\frac{M_{\rm BH}}{3M_{\odot}}\right)^{7/10}\left(\frac{r}{R_{g}}\right)^{9/5}
\label{eq:Mdotcond2}
\ee
Now, since most of the total neutrino luminosity is released from radii just
outside the innermost stable circular orbit $R_{\rm isco}$, we can evalulate
equation (\ref{eq:Mdotcond2}) at $r \approx 2R_{\rm isco}$ to define a critical
accretion rate
\be \dot{M}_{\rm opaque} \simeq 0.15M_{\odot}{\rm\,s^{-1}}\,\left(\frac{\alpha}{0.03}\right)^{4/5}\left(\frac{M_{\rm BH}}{3M_{\odot}}\right)^{7/10}\left(\frac{R_{\rm isco}}{6R_{g}}\right)^{9/5}
\label{eq:Mdotopaque}
\ee
above which the majority of the radiated energy will originate from an
optically-thick environment.  This expression agrees reasonably well with CB07,
who find that for $\alpha = 0.1$, $\dot{M}_{\rm opaque} = 0.7 M_{\odot}$
s$^{-1}$ and $0.06M_{\odot}$ s$^{-1}$ for a BH of spin $a = 0 (R_{\rm isco} =
6R_{g})$ and $a = 0.95 (R_{\rm isco} \simeq 1.5R_{g}$), respectively, although
the dependence on $\alpha$ that they find, $\dot{M}_{\rm opaque} \propto
\alpha$ is slightly different than than the $\propto \alpha^{4/5}$ dependence
that we find in equation (\ref{eq:Mdotopaque}).

\section{Self-Irradiation and Weak Interactions}
\label{s:weak_rates_appendix}

The implementation of charged-current weak interaction and self-irradiation follows
closely that of \citet{F12} in the context of core-collapse supernovae, with suitable 
modifications to account for the disk geometry. We consider only electron-type neutrinos 
and antineutrinos, with emission and absorption rates following \citet{bruenn85}.

\begin{figure}
\includegraphics*[width=\columnwidth]{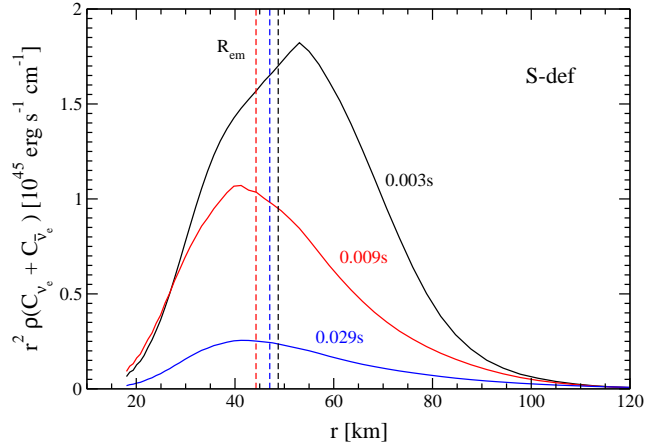}
\caption{Snapshots of the angle-integrated neutrino emissivity for model S-def. Times shown correspond
to orbits 1 (black), 3 (red), and 10 (blue) at $r=R_0$. The vertical dashed line shows the location
of $R_{\rm em}$ (eq.~[\ref{eq:R_em_def}]) in each of the models.}
\label{f:cooling_peak}
\end{figure}

Our prescription for self-irradiation assumes that neutrinos are emitted from a ring of material at 
a radius $R_{\rm em}$. This approximation is motivated by the fact that the neutrino
emissivity displays a single peak inside the disk, where most of the emission is
generated. Figure~\ref{f:cooling_peak} illustrates the shape of the function for model S-def.
We choose the emission radius to be an average weighted by the neutrino 
emissivity (eq.~[\ref{eq:R_em_def}]). This location coincides within $\sim 10\%$ with the radius inside which
$50\%$ of the neutrino emission is generated.

\begin{figure}
\includegraphics*[width=\columnwidth]{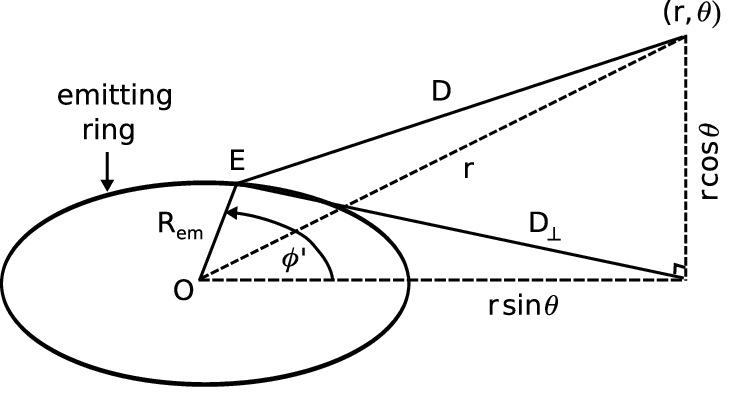}
\caption{Geometry for self-irradiation. All of the neutrino luminosity is assumed to originate
in a ring on the midplane with radius $R_{\rm em}$. A point in the disk with coordinates 
$(r,\theta)$ lies at a distance $D$ from an emitting element $E$ with azimuthal angle 
$\phi^\prime$. The total emission received by the point in the disk is
a sum over the emission from each ring element, leading to an angular dependence 
$I_{\rm ang}$ (eq.~[\ref{eq:int_angular}], Figure~\ref{f:int_angular}).}
\label{f:annular_geometry}
\end{figure}

Each fluid element on the ring is assumed to emit isotropically, with a total luminosity equal to the instantaneous
volume integral of the emissivities, and with an energy spectrum following a Fermi-Dirac distribution with zero 
chemical potential. 
The number density of neutrino species $i$ per unit 
energy $\epsilon$, propagation direction $\hat k$, and angular position on the ring $\phi^\prime$ is
\begin{eqnarray}
\frac{\partial^3 n_{\nu_i}}{\partial \phi^\prime\,\partial \cos\theta_k\,\partial \epsilon} & = & 
\frac{2\pi}{(hc)^3}\,f_{\nu_i}(\epsilon,T_{\nu_i},L_{\nu_i};r,\theta)\qquad\qquad\nonumber\\
& = &\frac{2\pi}{(hc)^3}\,
\frac{L_{\nu_i}/2\pi}{(7/16)4\pi \mathcal{R}^2\sigma_{\rm SB}T_{\nu_i}^4}\,\nonumber\\
&&\times F_{\rm FD}(\epsilon,T_{\nu_i},0)\,\Theta(\cos\theta_k - \cos\theta_{k,\rm min})\nonumber\\
\end{eqnarray}
where $\theta_k$ is the angle between the propagation vector and the radial direction,
and $\Theta$ is the step function. 
The intensty is assumed to be azimuthally symmetric around the propagation direction.
The Fermi-Dirac function is given by
\begin{equation}
F_{\rm FD}(\epsilon,T,\mu) = \frac{1}{e^{(\epsilon-\mu)/kT} + 1},
\end{equation}
and the minimum angle between $\mathbf k$ and the radial direction is
\begin{equation}
\label{eq:theta_k_min}
\cos\theta_{k,\rm min} \simeq 1 - \frac{1}{2}\left(\frac{\mathcal{R}}{D} \right)^2
\end{equation}
to lowest order in $\mathcal{R}/D$, 
where $\mathcal{R}$ is the radius of a spherical emitting element in the ring, and
\begin{eqnarray}
\label{eq:annular_distance}
D(r,\theta,R_{\rm em},\phi^\prime) = r\left[1 + \left(\frac{R_{\rm em}}{r}\right)^2 
-2\frac{R_{\rm em}}{r}\,\sin\theta\,\cos\phi^\prime\right]^{1/2}\nonumber\\
\label{eq:D_definition}
\phantom{A = B}
\end{eqnarray}
is the distance to a point in the disk with coordinates $(r,\theta)$ from 
the emitting blob in the ring, as depicted in Figure~\ref{f:annular_geometry}.
Keeping the lowest order terms arising from equation~(\ref{eq:theta_k_min}) ensures that
$\mathcal{R}$ scales out of the problem. By using equation~(\ref{eq:D_definition}) to capture
the global geometry of the problem, we are ignoring light bending and
Doppler effects. 

The rates per baryon of production and destruction of electrons and positrons 
that enter equation~(\ref{eq:gamma_net}) are given respectively by
\begin{eqnarray}
\label{eq:gamma_nem}
\Gamma_{e-} & = & \frac{2\pi m_{\rm n} c}{(hc)^3\rho}
		  \left[\int\totd\phi^\prime \int \totd\cos\theta_k
		  \int\epsilon^2 \totd\epsilon\,\left(\kappa_{\nu_e}+j_{\nu_e}\right)\, f_{\nu_e}\right.\nonumber\\
		  &&\left. \phantom{\frac{2\pi m_{\rm n} c}{(hc)^3\rho}} 
		    - \int \totd\cos\theta_k\int\epsilon^2 \totd\epsilon j_{\nu_e}\right]\\
\label{eq:gamma_nep}
\Gamma_{e+} & = &\frac{2\pi m_{\rm n} c}{(hc)^3\rho}
		  \left[\int\totd\phi^\prime \int \totd\cos\theta_k 
		  \int \epsilon^2 \totd\epsilon\,\left(\kappa_{\bar\nu_e} + j_{\bar\nu_e} \right)\,f_{\bar\nu_e}\right.\nonumber\\
		  &&\left. \phantom{\frac{2\pi m_{\rm n} c}{(hc)^3\rho}}
		    - \int \totd\cos\theta_k \int \epsilon^2 \totd\epsilon j_{\bar\nu_e}\right]
\end{eqnarray}
where $j_{\nu_i}$ and $\kappa_{\nu_i}$ are the emissivity and absorption coefficient, respectively, associated with
electron-type neutrinos or antineutrinos (as subscripted). The coefficients are given by \citep{bruenn85}
\begin{eqnarray}
\label{eq:jnu_em}
j_{\nu_e}(\epsilon,T,\mu_e,n_p) & = & \frac{\tilde{G}_{\rm F}^2}{\pi}\left(g_{\rm V}^2 + 3g_{\rm A}^2\right)\, n_p
                                    F_{\rm FD}(\epsilon,T,\mu_e)\nonumber\\
				    &&\times\left[\epsilon + \Delta_m \right]^2
                                    \left[1 - \frac{m_e^2 c^4}{(\epsilon + \Delta_m)^2}\right]^{1/2}\\
\label{eq:jnu_ep}
j_{\bar{\nu_e}}(\epsilon,T,\mu_e,n_p) & = & \frac{\tilde{G}_{\rm F}^2}{\pi}\left(g_{\rm V}^2 + 3g_{\rm A}^2\right)\, n_n\nonumber\\
                                    &&\times F_{\rm FD}(\epsilon-\Delta_m,T,-\mu_e)\nonumber\\
				    &&\times\left[\epsilon - \Delta_m \right]^2
                                    \left[1 - \frac{m_e^2 c^4}{(\epsilon - \Delta_m)^2}\right]^{1/2}\nonumber\\
				    &&\times\Theta(\epsilon-\Delta_m - m_e c^2)\\
\label{eq:knu_em}
\kappa_{\nu_e}(\epsilon,T,\mu_e,n_p) & = & \frac{\tilde{G}_{\rm F}^2}{\pi}\left(g_{\rm V}^2 + 3g_{\rm A}^2\right)\, n_n
                                    \left[1-F_{\rm FD}(\epsilon,T,\mu_e)\right]\nonumber\\
				    && \times\left[\epsilon + \Delta_m \right]^2
                                    \left[1 - \frac{m_e^2 c^4}{(\epsilon + \Delta_m)^2}\right]^{1/2}\\
\label{eq:knu_ep}
\kappa_{\bar{\nu_e}}(\epsilon,T,\mu_e,n_p) & = & \frac{\tilde{G}_{\rm F}^2}{\pi}\left(g_{\rm V}^2 + 3g_{\rm A}^2\right)\, n_p\nonumber\\
                                    &&\times\left[1-F_{\rm FD}(\epsilon-\Delta_m,T,-\mu_e)\right]\nonumber\\
				    &&\times\left[\epsilon - \Delta_m \right]^2
                                    \left[1 - \frac{m_e^2 c^4}{(\epsilon - \Delta_m)^2}\right]^{1/2}\nonumber\\
                                    &&\times\Theta(\epsilon-\Delta_m - m_e c^2),
\end{eqnarray}
where $\mu_e$ is the chemical potential of electrons, $n_n$ and $n_p$ the number density of free neutrons and protons, respectively,
$\tilde{G}_{\rm F} = G_F/(\hbar c)^3$ the Fermi constant, $g_{\rm V}$ and $g_{\rm A}$ the vector
and axial coupling constants, respectively, and $\Delta_m = (m_n - m_p) c^2$ the difference between the
rest mass energy of neutrons and protons.

The energy source terms that enter equation~(\ref{eq:Q_net}) are
\begin{eqnarray}
\label{eq:H_em}
\mathcal{H}_{\nu_e}     & = &  \frac{2\pi m_{\rm n} c}{(hc)^3\rho}\int\totd\phi^\prime\,
				\int \totd\,\cos\theta_k \int \epsilon^3 \totd\,\epsilon
                               \left[ j_{\nu_e} + \kappa_{\nu_e}\right]f_{\nu_e}\nonumber\\
			&&\\
\label{eq:H_ep}
\mathcal{H}_{\bar\nu_e} & = &  \frac{2\pi m_{\rm n} c}{(hc)^3\rho}\int\totd\phi^\prime
				\int \totd\,\cos\theta_k \int \epsilon^3 \totd\,\epsilon
                       \left[ j_{\bar\nu_e} + \kappa_{\bar\nu_e}\right]f_{\bar\nu_e}\nonumber\\
			&&\\
\label{eq:C_em}
\mathcal{C}_{\nu_e}     & = &  \frac{4\pi m_{\rm n} c}{(hc)^3\rho}\int \epsilon^3 \totd\,\epsilon \, j_{\nu_e}\\
\label{eq:C_ep}
\mathcal{C}_{\bar\nu_e} & = &  \frac{4\pi m_{\rm n} c}{(hc)^3\rho}\int \epsilon^3 \totd\,\epsilon \, j_{\bar\nu_e}.
\end{eqnarray}

\begin{figure}
\includegraphics*[width=\columnwidth]{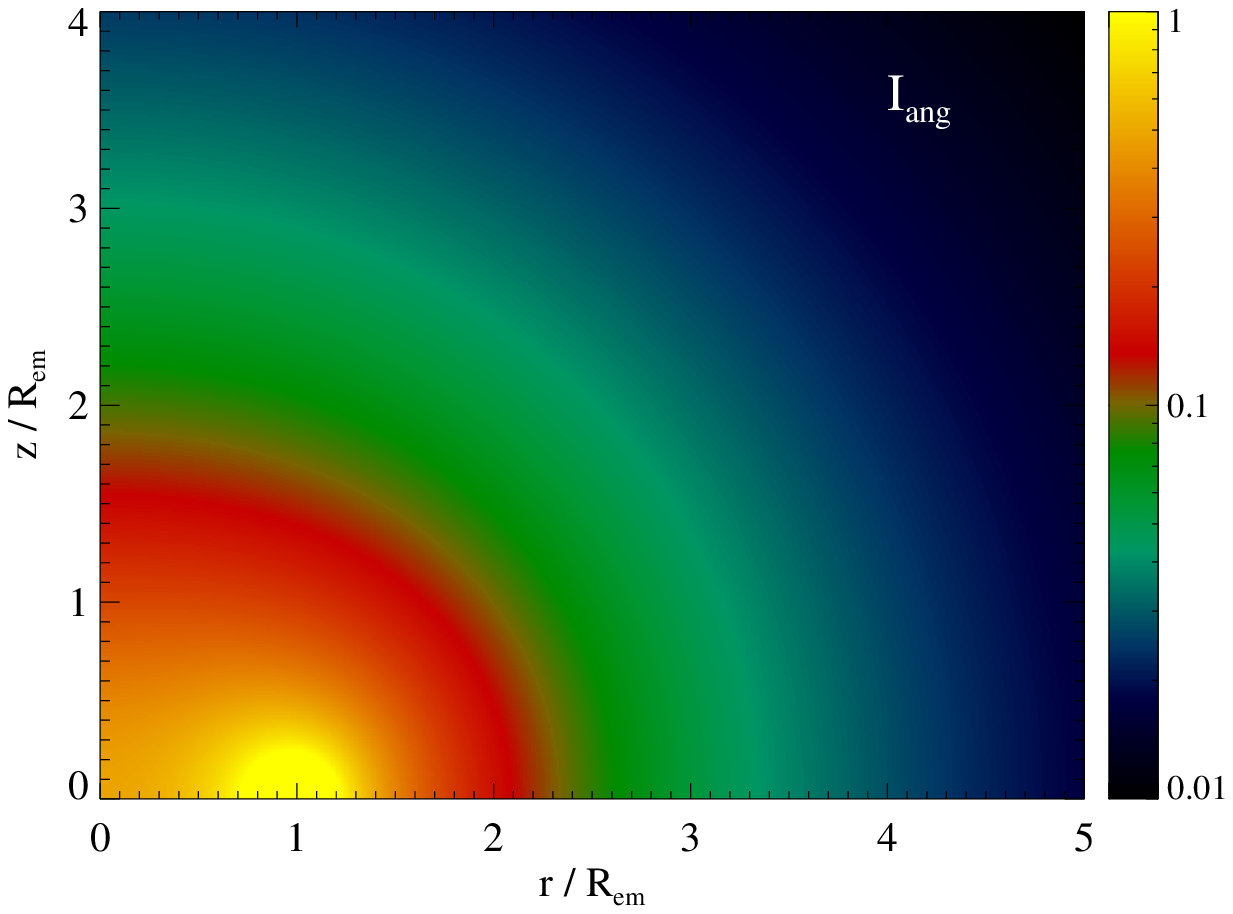}
\caption{Angular dependence of the radiation field (eq.~[\ref{eq:int_angular}]).
The angular factor approaches a pure inverse square law at large distances from 
the origin (eq.~[\ref{eq:int_angular_approx}]).}
\label{f:int_angular}
\end{figure}

The anisotropy of the global radiation field decouples from the energy integrals,
and enters through a dimensionless prefactor
\begin{eqnarray}
\label{eq:int_angular}
I_{\rm ang} & = & \frac{1}{2\pi} \left(\frac{R_{\rm em}}{r} \right)^2        
                  \int_0^{2\pi}\frac{\totd \phi^\prime}{2\left[D(r,\theta,R_{\rm em},\phi^\prime)/r \right]^2}\\
\label{eq:int_angular_approx}
 & \simeq & \frac{1}{2}\left(\frac{R_{\rm em}}{r} \right)^2\left[1 -\left(\frac{R_{\rm em}}{r}\right)^2\right] 
           + \frac{1}{2\pi}\left(\frac{R_{\rm em}}{r} \right)^4\sin^2\theta\nonumber\\
\end{eqnarray}
in all terms containing an $f_{\nu_i}$ factor ($D$ is given by eq.~[\ref{eq:annular_distance}]).
The spatial dependence of $I_{\rm ang}$ is shown in Figure~\ref{f:int_angular}. In the inner
regions of the disk, this function varies by factors of order unity following the annular geometry
of the emission region. It becomes increasingly spherical for larger radii.
Quantitatively, this can be seen from
equation~(\ref{eq:int_angular_approx}), which is a series expansion in $R_{\rm em}/r$. Due to
the azimuthal averaging, the leading error term scales like the 4th power of this ratio.
The integral diverges when $r=R_{\rm em}$. To avoid excessive heating by artificial concentration
of the emission at this radius, and given that the spatial distribution of the emission has 
a finite width (Figure~\ref{f:cooling_peak}), we impose $I_{\rm ang}\leq 1$. The effect of this 
upper limit can also be seen in Figure~\ref{f:int_angular}.

All terms containing a $f_{\nu_i}$ factor are also multiplied by a normalization
constant $N_{\nu_i}$
\begin{equation}
N_{\nu_i} = \frac{L_{\nu_i}}{(7/16)4\pi R_{\rm em}^2 \sigma_{\rm SB}T_{\nu}^4},
\end{equation}
which ensures that the incident flux is consistent with the total neutrino
luminosity $L_{\nu_i}$ emitted by the disk. The denominator normalizes the assumed 
Fermi-Dirac spectrum in $f_{\nu_i}$. The neutrino temperature used is the same
for both neutrino species, and is equal to the
average, mass-weighted gas temperature in an (angle integrated) annular region 
centered on $R_{\rm em}$, with a radial width of $10\%$. This averaging is performed
to smooth over stochastic fluctuations in localized disk regions.

Weak rates are tabulated as a function of the gas temperature $T$, electron
degeneracy parameter including rest mass $\mu_e/(k_{\rm B}T)$, and
neutrino temperature $T_\nu$. The angular function $I_{\rm ang}$ is also
tabulated in cylindrical coordinates, covering the range $(r/R_{\rm em}) \in [0,10]$.
Outside of this range, the asymptotic expansion in equation~(\ref{eq:int_angular_approx}) is used.

\section{Heavy Element Nucleosynthesis}
\label{s:nucleo}

Here we review the conditions for nucleosynthesis of heavy elements in hot outflows, such as those from binary NS merger accretion disks.  Free nuclei recombine into $\alpha$-particles once the temperature decreases to $T \lesssim 10^{10}$ K, the energy from which plays a key role in powering the disk outflows explored in this paper.  Heavier elements start to form once the temperature decreases further, $T \lesssim 5\times 10^{9}$ K, such that the reaction $^{4}$He($\alpha$n,$\gamma$)$^{9}$Be($\alpha$,n)$^{12}$C occurs (at higher temperatures, the abundance of $^{9}$Be is limited by photodisintegration).  

Immediately after $^{12}$C forms, multiple additional $\alpha$-captures produce heavy `seed' nuclei with characteristic mass $\bar{A} \simeq 90-120$ and charge $\bar{Z}\simeq 35$ (the `$\alpha$-process'; \citealt{Woosley&Hoffman92}).  Whether nucleosynthesis proceeds to even heavier $r$-process nuclei depends primarily on the ratio of free neutrons to seed nuclei once the $\alpha$-process completes.  Since the formation of $^{12}$C is the rate limiting step in forming seeds, the quantity and distribution of $r$-nuclei nuclei synthesized depends on the entropy $S$, electron fraction $Y_e$, and expansion timescale $t_{\rm exp}$ at the radius where $T \simeq 5\times 10^{9}$ K.

 If the electron fraction of the outflow is less than that of the seed nuclei themelves, i.e.~$Y_e \lesssim \frac{\bar{Z}}{\bar{A}} \simeq 0.3$, then the $^{4}$He($\alpha$n,$\gamma$)$^{9}$Be($\alpha$,n)$^{12}$C reaction is ultimately limited by the total number of $\alpha$-particles (the neutron abundance can be considered fixed for purposes of calculating the $^{12}$C reaction yield).  In this case the final mass fraction of alpha particles is approximately given by \citep{Hoffman+97}
\be
X_{\rm He} = \left[(2Y_e)^{-2} + \mathcal{F}\right]^{-1/2},
\label{eq:XHe}
\ee 
where
\be
\mathcal{F} =1.8\times 10^{4}(1-2Y_e)\left(\frac{\bar{Z}}{36}\right)\left(\frac{t_{\rm exp}}{\rm 0.1\,s}\right)\left(\frac{S}{\rm 20 k_b\,nuc^{-1}}\right)^{-3},
\label{eq:F}
\ee
where $\mathcal{F}$ is proportional to the reaction rate for $^{4}$He $\rightarrow$ $^{12}$C integrated over the timescale available for burning $\sim t_{\rm exp}$.  The reaction rate depends on $S^{-3}$ since $\rho \propto 1/S$ at fixed temperature, and since $^{4}$He($\alpha$n,$\gamma$)$^{9}$Be($\alpha$,n)$^{12}$C is an effective four-body reaction.  Any mass not trapped in $\alpha$-particles ends up in $r$-process nuclei with mass fraction 
\be
X_{\rm r} = 1-X_{\rm He},
\ee
and mass number greater than
\be
A = \bar{A}\left(1 - \frac{1-2Y_e}{X_r}\right)^{-1}.
\ee

If $\mathcal{F} \ll 1$ then very little $^{12}$C forms (`He-rich freeze-out'), in which case helium achieves its maximum abundance $X_{\rm He} = 2Y_e$, unchanged from its value just after $\alpha$ recombination.  Since most of the available protons are trapped in $^{4}$He, the $r$-process nuclei produced in this case are very massive, easily reaching beyond the third peak $A \gtrsim 195$ (and likely undergoing fission).  Even if $\mathcal{F} = \infty$ ($X_{\rm r} = 1$), however, nucleosynthesis will still reach the Lanthanide elements ($A \gtrsim $ 139 = $A_{\rm La}$) if the outflow is sufficiently neutron-rich: $Y_{e} \lesssim \frac{\bar{A}}{2A_{\rm La}} \approx 0.30-0.40$.  Thus, given the low electron fraction of the disk outflows that we find in $\S\ref{s:outflows}$ ($\bar{Y}_e \approx 0.2$), we expect Lathanides to easily form, resulting in a high optical opacity (\citealt{Kasen+13}).

\bibliographystyle{mn2e}
\bibliography{ms}

\label{lastpage}
\end{document}